\documentclass[prd,preprint,superscriptaddress,showpacs,byrevtex]{revtex4}
\usepackage{bm}
\def\fsl#1{\setbox0=\hbox{$#1$}           % set a box for #1
   \dimen0=\wd0                                 % and get its size
   \setbox1=\hbox{/} \dimen1=\wd1               % get size of /
   \ifdim\dimen0>\dimen1                        % #1 is bigger
      \rlap{\hbox to \dimen0{\hfil/\hfil}}      % so center / in box
      #1                                        % and print #1
   \else                                        % / is bigger
      \rlap{\hbox to \dimen1{\hfil$#1$\hfil}}   % so center #1
      /                                         % and print /
   \fi}                                         %
\newcommand{\be}{\begin{equation}}
\newcommand{\ee}{\end{equation}}
\newcommand{\bea}{\begin{eqnarray}}
\newcommand{\eea}{\end{eqnarray}}
\newcommand{\beq}{\begin{equation}}
\newcommand{\eeq}{\end{equation}}
\newcommand{\beqs}{\begin{eqnarray}}
\newcommand{\eeqs}{\end{eqnarray}}

\begin{document}
\title{ Correct Definition of the Gluon Distribution Function at High Energy Colliders }
\author{Gouranga C Nayak }\email{nayak@max2.physics.sunysb.edu}
\affiliation{ 665 East Pine Street, Long Beach, New York 11561, USA }
\date{\today}
\begin{abstract}
Unlike QED, since $F_{\mu \nu}^a(x)F^{\mu \nu a}(0)$ in QCD contains cubic and quartic powers of the gluon field the present
definition of the gluon distribution function at high energy colliders is not consistent with the number operator
interpretation of the gluon. In this paper we derive the correct definition of the gluon distribution function
at high energy colliders from first principles which is consistent with the number operator interpretation of the gluon and is
gauge invariant and is consistent with the factorization theorem in QCD.
\end{abstract}
\pacs{12.38.Lg; 12.38.Aw; 14.70.Dj; 12.39.St }
\maketitle
\pagestyle{plain}
\pagenumbering{arabic}
\section{Introduction}
Quark and gluon distribution functions (parton distribution functions (PDFs)) inside the hadron play a significant role to study
standard model and beyond standard model physics at high energy colliders. Consider for example, the Higgs  \cite{higgs, higgs1} discovery
at large hadron colliders (LHC) \cite{atlas,cms}. The gluon-gluon fusion process ($gg \rightarrow H$) contribute significantly
to the Higgs production cross section at LHC \cite{gg}. Similarly, the gluon fusion and
gluon fragmentation processes contribute dominantly to the production of the heavy quark, heavy quarkonium and jet etc.
at high energy colliders. Hence, as the hadron collider's total energy becomes higher and higher, the gluon distribution
function inside the hadron plays
significant role to study standard model and beyond standard model physics at high energy colliders.

Unfortunately, the parton distribution function inside a hadron is not calculated so far because it is a non-perturbative quantity
in QCD and we do not know the solution of the non-perturbative QCD yet.
Hence the usual procedure in the high energy phenomenology is to determine the parton distribution function
from a set of experiments at some momentum transfer scale and then determine its evolution to other
momentum transfer scale by using DGLAP evolution equations \cite{dglap}. Once the parton distribution function
is determined in this way then one uses it to predict physical observable at other collider experiments.
Hence in order for this prescription to work it is necessary that these parton distribution functions
are universal, {\it i. e.}, they do not change from one collider experiment to another
collider experiment. From this point of view it is necessary to use the correct definition of the parton distribution function at
high energy colliders. If one does not use the correct definition of the parton distribution
function at high energy colliders then one will obtain divergent cross section for physical
observable at high energy colliders. This can be seen as follows.

Consider, for example, the hadron production at high energy colliders. If the factorization theorem holds
\cite{sterman,collins,cs2,bodwin} then the formula for the hadron production cross section at high energy hadronic colliders is given by
\bea
\sigma (AB \rightarrow H+X) ~=~ \sum_{a,b}~\int dx_1 ~\int dx_2 ~\int dz~f_{a/A}(x_1,Q^2)~f_{b/B}(x_2,Q^2)~{\hat \sigma}(ab \rightarrow cd)~D_{H/c}(z,Q^2) \nonumber \\
\label{cr1}
\eea
where $f_{a/A}(x,Q^2)$ is the parton distribution function (PDF), ${\hat \sigma}(ab \rightarrow cd)$ is the partonic level scattering
cross section and $D_{H/c}(z,Q^2)$ is the fragmentation function (FF). For non-hadronic observable at high energy colliders the
fragmentation function $D_{H/c}(z,Q^2)$ is not included in eq. (\ref{cr1}). In the above equation $a,b,c,d=q,{\bar q},g$ where $q$,
${\bar q}$ and $g$ are quark, antiquark and gluon.

Hence if one does not use the correct definition of the parton distribution function or fragmentation function
in eq. (\ref{cr1}) then any uncanceled collinear or soft divergences in the Feynman diagrams in the
partonic level scattering cross section ${\hat \sigma}(ab \rightarrow cd)$ can not be absorbed in the definition of the non-perturbative
parton distribution function $f_{a/H}(x,Q^2)$ or fragmentation function $D_{H/a}(x,Q^2)$ in eq. (\ref{cr1}). Hence it is necessary to
derive the correct definition of the parton distribution function $f_{a/H}(x,Q^2)$ and fragmentation function $D_{H/a}(x,Q^2)$ from
first principles.

A correct definition of a parton distribution function at high energy colliders must satisfy the following three properties: 1)
it must be consistent with the number operator interpretation in quantum mechanics, {\it i.e.},
$\psi^\dagger(x)\psi(0)$ or $F^{\mu \nu a}(x)F_{\mu \nu}^a(0)$
must be proportional to quadratic power of the quark or (quantum) gluon field, 2) it must be gauge invariant and 3)
it must be consistent with factorization theorem in QCD. The present definition of the quark distribution function
at high energy colliders is consistent with these three properties \cite{collins}. However, the present definition of the gluon
distribution function at high energy colliders \cite{collins} is not consistent with these three properties as we will
see below.

The present definition of the gluon distribution function at high energy colliders is given by \cite{collins}
\bea
&& f_{g/P}(x)
= \frac{1}{2\pi xP^+}~\int dy^- e^{-ix{P}^+ y^-} \nonumber \\
&&[<P| F^{+\mu a}(0,y^-,0_T) ~\times ~ {\cal P}{\rm exp}[igT^{(A)b}\int_0^{y^-}dz^-Q^{+b}(0,z^-,0_T)]~\times ~F_\mu^{~+a}(0)  |P>]
\label{gpdf}
\eea
where
\bea
F_{\mu \nu}^a(x)=\partial_\mu Q_\nu^a(x)-\partial_\nu Q_\mu^a(x)+gf^{abc}Q_\mu^b(x)Q_\nu^c(x),
\label{ffmnx}
\eea
$T^{(A)a}_{bc}=-if^{abc}$ and $Q^{\mu a}(x)$ is the quantum gluon field.

The expression ${\cal P}{\rm exp}[igT^{(A)b}\int_0^{y^-}dz^-Q^{+b}(0,z^-,0_T)]$ in eq. (\ref{gpdf})
stands for the light-like Wilson line. Since the light-like Wilson line in QCD involves
soft-collinear gluon field $A^{\mu a}(x)$, it can be treated classically as $A^{\mu a}(x)$ becomes
the SU(3) pure gauge when Wilson line becomes light-like
(see section II for details). Hence for the better understanding of the number operator interpretation
of the gluon in eq. (\ref{gpdf}) let us replace the light-like Wilson line
${\cal P}{\rm exp}[igT^{(A)b}\int_0^{y^-}dz^-Q^{+b}(0,z^-,0_T)]$ in eq. (\ref{gpdf})
by its classical counterpart ${\cal P}{\rm exp}[igT^{(A)b}\int_0^{y^-}dz^-A^{+b}(0,z^-,0_T)]$
where $A^{\mu a}(x)$ is the classical SU(3) pure gauge background field.
With this one finds from eq. (\ref{ffmnx}) that $F^{\mu \nu a}(x)F_{\mu \nu}^a(0)$
in eq. (\ref{gpdf}) contains cubic and quartic powers of the (quantum) gluon field $Q^{\mu a}(x)$ which implies that the
present definition of gluon distribution function at high energy colliders
in eq. (\ref{gpdf}) is not consistent with the number operator interpretation
in quantum mechanics, {\it i.e.}, $F^{\mu \nu a}(x)F_{\mu \nu}^a(0)$
is not proportional to quadratic power of the (quantum) gluon field $Q^{\mu a}(x)$.
This implies that the present definition of the gluon distribution function at high energy colliders is not correct.

In this paper we derive correct definition of the gluon distribution function at high energy colliders from first
principles which is consistent with the number operator interpretation of the gluon and is gauge invariant and is
consistent with the factorization theorem in QCD.

We find that the correct definition of the gluon distribution function at high energy colliders which is
consistent with the number operator interpretation of the gluon and is gauge invariant and is consistent with the
factorization theorem in QCD is given by
\bea
&& f_{g/P}(x)= \frac{P^+}{x2\pi}\int dy^- e^{-ix{P}^+ y^- }\times <P| Q_\mu^a (0,y^-,0_T)
[{\cal P}{\rm exp}[igT^{(A)b}\int_0^{y^-}dz^-A^{+b}(0,z^-,0_T)]]Q^{\mu a}(0)|P>\nonumber \\
\label{gpdffpi}
\eea
which is valid in covariant gauge, in light-cone gauge, in general axial gauges, in general non-covariant gauges and in
general Coulomb gauge etc. respectively. Note that the correct definition of gluon distribution function
at high energy colliders in eq. (\ref{gpdffpi}) is gauge invariant with respect to the
gauge transformation
\bea
T^aA'^a_\mu(x) = U(x)T^aA^a_\mu(x) U^{-1}(x)+\frac{1}{ig}[\partial_\mu U(x)] U^{-1}(x),~~~~~~~~~~~U(x)=e^{igT^a\omega^a(x)}
\label{aftgrmpi}
\eea
along with the homogeneous transformation \cite{thooft,zuber,zuber1,abbott}
\bea
T^aQ'^a_\mu(x)=U(x)T^aQ^a_\mu(x) U^{-1}(x),~~~~~~~~~~~U(x)=e^{igT^a\omega^a(x)}
\label{jprtpi}
\eea
where $Q^{\mu a}(x)$ is the hard (quantum) gluon field whose distribution function we measure
and $A^{\mu a}(x)$ is the soft-collinear gluon field which is the SU(3) pure gauge background field.
After renormalization the gluon distribution function
is expected to obey a QCD evolution equation, like DGLAP equation \cite{tung}, which follows from renormalization group
equation.

We will provide a derivation of eq. (\ref{gpdffpi}) in this paper.

The paper is organized as follows. In section II we describe soft-collinear divergences and light-like Wilson
line in QCD. In section III we discuss the non-perturbative gluon correlation function and proof of
factorization theorem in covariant gauge. In sections IV, V, VI and VII we discuss the non-perturbative gluon correlation
function and proof of factorization theorem in general axial gauges, in light-cone gauge, in general non-covariant gauges and in
general Coulomb gauge respectively. In section VIII we derive the correct definition of the gluon distribution
function at high energy colliders as given by eq. (\ref{gpdffpi}). Section IX contains conclusions.

\section{ Soft-Collinear Divergences and Light-Like Wilson Line in QCD }

Note that in eq. (\ref{cr1}) the collinear and soft divergences can occur in the Feynman diagrams in partonic level cross sections.
Hence it is important to show that these collinear/soft divergences are factorized into the definition of the parton distribution
function/fragmentation function. This is done by supplying Wilson line in the definition of parton distribution
function/fragmentation function which also makes them gauge invariant. Hence we will briefly discuss the issue of gauge invariance
and the Wilson line for soft-collinear divergences in QCD in this section.

Before proceeding to the issue of gauge invariance and the Wilson line for soft-collinear divergences in QCD let us first discuss
the corresponding situation in QED. The gauge transformation of the Dirac field of the electron in QED is given by
\bea
\psi'(x)=e^{ie\omega(x)}\psi(x).
\label{phg3q}
\eea
Hence we can expect to address the issue of gauge invariance and factorization of soft-collinear divergences in QED simultaneously
if we can relate the $\omega(x)$ to the photon field $A^\mu(x)$. This is easily done by using Eikonal Feynman rules
in QED for soft-collinear divergences which can be seen as follows.

The Eikonal propagator times the Eikonal vertex for a photon with four momentum $k^\mu$ interacting with an
electron moving with four momentum $p^\mu$ in the limit $|{\vec k}|<<< |{\vec p}|$ is given by
\cite{collins,tucci,collinssterman,berger,frederix,nayakqed,scet1,pathorder,nayaka2,nayaka3}
\bea
e~\frac{p^\mu}{p \cdot k+i\epsilon }.
\label{eikonaliqp}
\eea
In QED the soft divergences arise only from the
emission of a photon for which all components of the four-momentum $k^\mu$ are small
($|{\vec k}| \rightarrow 0$) which is evident from eq. (\ref{eikonaliqp}). From eq. (\ref{eikonaliqp})
one also finds that when $0<|{\vec k}|<<< |{\vec p}|$ and ${\vec p}$ is parallel to ${\vec k}$ one may
find collinear divergences.

However, since
\bea
p \cdot k =p_0 k_0 -{\vec p} \cdot {\vec k}=|{\vec p}||{\vec k}|(\sqrt{1+\frac{m^2}{{\vec p}^2}}  -{\rm cos}\theta)
\label{scd}
\eea
one finds that the collinear divergences does not appear in QED because of the non-vanishing mass of the electron,
{\it i. e.}, $m\neq 0$. From eqs. (\ref{eikonaliqp}) and (\ref{scd}) one finds that the collinear divergences appear only
when $m=0$ and $\theta =0$ where $\theta$ is the angle between ${\vec p}$ and ${\vec k}$.
Since gluons are massless and the massless gluons interact with each other one finds that the collinear
divergences appear in QCD. Since a massless particle is always light-like one
finds that the soft-collinear divergences can be described by light-like Wilson line in QCD.

For light-like electron we find from eq. (\ref{eikonaliqp})
\bea
e~\frac{p^\mu}{p \cdot k+i\epsilon }=e~\frac{l^\mu}{l \cdot k+i\epsilon }
\label{eikonaliq}
\eea
where $l^\mu$ is the light-like four-velocity ($|{\vec l}|$=1) of the electron.
Note that when we say the "light-like electron" we mean the electron
that is traveling at its highest speed which is arbitrarily close to the speed of light
($|{\vec l}|\sim 1$) as it can not travel exactly at speed of light
because it has finite mass even if the mass of the electron is very small. Hence we find that
if $l^\mu$ is light-like four velocity then the soft-collinear divergences can be described by
the Eikonal Feynman rule as given by eq. (\ref{eikonaliq}).

From eq. (\ref{eikonaliq}) we find
\bea
&& e\int \frac{d^4k}{(2\pi)^4} \frac{l\cdot {  A}(k)}{l\cdot k +i\epsilon } =-e i\int_0^{\infty} d\lambda \int \frac{d^4k}{(2\pi)^4} e^{i l \cdot k \lambda} l\cdot {A}(k) = ie\int_0^{\infty} d\lambda l\cdot { A}(l\lambda)
\label{ftgtq}
\eea
where the photon field  $ { A}^{\mu }(x)$ and its Fourier transform $ { A}^{\mu }(k)$ are related by
\bea
{ A}^{\mu }(x) =\int \frac{d^4k}{(2\pi)^4} { A}^{\mu }(k) e^{ik \cdot x}.
\label{ftq}
\eea
Now consider the corresponding Feynman diagram for the soft-collinear divergences in QED
due to exchange of two soft-collinear photons of four-momenta $k^\mu_1$ and $k^\mu_2$.
The corresponding Eikonal contribution due to two soft-collinear photons exchange is analogously given by
\bea
&& e^2\int \frac{d^4k_1}{(2\pi)^4} \frac{d^4k_2}{(2\pi)^4} \frac{ l\cdot { A}(k_2) l\cdot { A}(k_1)}{(l\cdot (k_1+k_2) +i \epsilon)(l\cdot k_1 +i \epsilon)} \nonumber \\
&&=e^2i^2 \int_0^{\infty} d\lambda_2 \int_{\lambda_2}^{\infty} d\lambda_1 l\cdot { A}(l\lambda_2) l\cdot {\cal A}(l\lambda_1)
\nonumber \\
&&= \frac{e^2i^2}{2!}\int_0^{\infty} d\lambda_2 \int_0^{\infty} d\lambda_1 l\cdot {\cal A}(l\lambda_2) l\cdot {\cal A}(l\lambda_1).
\eea
Extending this calculation up to infinite number of soft-collinear photons
we find that the Eikonal contribution for the soft-collinear divergences due to
soft-collinear photons exchange with the light-like electron in QED is given by the exponential
\bea
e^{ie \int_0^{\infty} d\lambda l\cdot { A}(l\lambda) }
\label{iiijqed}
\eea
where $l^\mu$ is the light-like four velocity of the electron. The Wilson line in QED is given by
\bea
e^{ie \int_{x_i}^{x_f} dx^\mu A_\mu(x) }.
\label{klj}
\eea
When $A^\mu(x)=A^\mu(l\lambda)$ as in eq. (\ref{iiijqed}) then
one finds from eq. (\ref{klj}) that the light-like Wilson line in QED for soft-collinear divergences is given by \cite{stermanpath}
\bea
e^{ie \int_0^x dx^\mu A_\mu(x) }=e^{-ie \int_0^{\infty} d\lambda l\cdot { A}(x+l\lambda) }e^{ie \int_0^{\infty} d\lambda l\cdot { A}(l\lambda) }.
\label{tto}
\eea
Note that a light-like electron traveling with light-like four-velocity $l^\mu$ produces U(1) pure gauge potential $A^{\mu }(x)$
at all the time-space position $x^\mu$ except at the position ${\vec x}$ perpendicular to the direction of motion
of the electron (${\vec l}\cdot {\vec x}=0$) at the time of closest approach \cite{collinssterman,nayakj,nayake}.
When $A^{\mu }(x) = A^{\mu }(\lambda l)$ as in eq. (\ref{iiijqed})
we find ${\vec l}\cdot {\vec x}=\lambda {\vec l}\cdot {\vec l}=\lambda\neq 0$ which implies that the light-like Wilson line
finds the photon field $A^{\mu }(x)$ in eq. (\ref{iiijqed}) as the U(1) pure gauge. The U(1) pure gauge is given by
\bea
A^\mu(x)=\partial^\mu \omega(x)
\label{purea1}
\eea
which gives from eq. (\ref{tto}) the light-like Wilson line in QED for soft-collinear divergences
\bea
e^{ie\omega(x)}e^{-ie\omega(0)}=e^{ie \int_0^x dx^\mu A_\mu(x) }=e^{-ie \int_0^{\infty} d\lambda l\cdot { A}(x+l\lambda) }e^{ie \int_0^{\infty} d\lambda l\cdot { A}(l\lambda) }
\label{lkj}
\eea
which depends only on end points $0$ and $x^\mu$ but is independent of the path. The path independence can also be found from
Stokes theorem because for pure gauge
\bea
F^{\mu \nu}(x)=\partial^\mu A^\nu(x)-\partial^\nu A^\mu(x)=0
\eea
which gives from Stokes theorem
\bea
e^{ie \oint_C dx^\mu A_\mu(x) }=e^{ie \int_S dx^\mu dx^\nu F_{\mu \nu}(x) }=1
\eea
where $C$ is a closed path and $S$ is the surface enclosing $C$. Now considering two different paths
$L$ and $M$ with common end points $0$ and $x^\mu$ we find
\bea
e^{ie \oint_C dx^\mu A_\mu(x) }=e^{ie \int_L dx^\mu A_\mu(x)- ie \int_M dx^\mu A_\mu(x)}=1
\eea
which implies that
\bea
e^{ie \int_0^x dx^\mu A_\mu(x) }
\eea
depends only on end points $0$ and $x^\mu$ but is independent of path which can also be seen from eq. (\ref{lkj}).
Hence from eq. (\ref{lkj}) we find that the abelian phase or the gauge link in QED is given by
\bea
e^{-ie \int_0^{\infty} d\lambda l\cdot { A}(x+l\lambda) }=e^{ie\omega(x)}.
\label{phas}
\eea
From eqs. (\ref{phg3q}) and (\ref{phas}) one expects that the gauge invariance and factorization of soft-collinear divergences
in QED can be explained simultaneously.

One can recall that the gauge invariant greens function in QED
\bea
G(x_1,x_2)=<{\bar \psi}(x_2)~\times~{\rm exp}[ie \int_{x_1}^{x_2} dx^\mu A_\mu(x)]~\times~ \psi(x_1)>
\label{qed}
\eea
in the presence of background field $A^\mu(x)$ was formulated by Schwinger long time ago \cite{schw1}.
When this background field $A^\mu(x)$ is replaced by the U(1) pure gauge background field
as given by eq. (\ref{purea1}) then one finds by using the path integral method of QED that \cite{tucci}
\bea
&&e^{ie\omega(x_2)}<{\bar \psi}(x_2)~ \psi(x_1)>_A e^{-ie\omega(x_1)}= <{\bar \psi}(x_2)~ \psi(x_1)>\nonumber \\
&&=e^{-ie\int_0^{\infty} d\lambda l\cdot { A}(x_2+l\lambda)}<{\bar \psi}(x_2)~ \psi(x_1)>_A e^{ie\int_0^{\infty} d\lambda l\cdot { A}(x_1+l\lambda)}
\label{tucci}
\eea
which proves the gauge invariance and factorization
of soft-collinear divergences in QED simultaneously.
In eq. (\ref{tucci}) the $<{\bar \psi}(x_2)~ \psi(x_1)>$ is the full Green's function
in QED and $<{\bar \psi}(x_2)~ \psi(x_1)>_A$ is the corresponding Green's function in
the background field method of QED. This path integral technique is also used in \cite{nayakqed} to prove factorization of
soft-collinear divergences in non-equilibrium QED.

Hence we find that the gauge invariance and factorization of soft-collinear divergences in QED can be studied by using
path integral method of QED in the presence of U(1) pure gauge background field.
Therefore one expects that the gauge invariance and factorization of soft-collinear divergences in QCD can be studied by using
path integral method of QCD in the presence of SU(3) pure gauge background field.

Now let us proceed to QCD.
The gauge transformation of the quark field in QCD is given by
\bea
\psi'(x)=e^{igT^a\omega^a(x)}\psi(x).
\label{phg3}
\eea
Hence one finds that the issue of gauge invariance and factorization of soft-collinear divergences in QCD can be
simultaneously explained if $\omega^a(x)$ can be related to the gluon field $ {A}^{\mu a}(x)$. This is
easily done by using Eikonal Feynman rules in QCD for soft-collinear divergences which can be seen as follows.

The Eikonal propagator times the Eikonal vertex for a gluon with four momentum $k^\mu$ interacting with a
quark moving with four momentum $p^\mu$ in the limit $|{\vec k}|<<< |{\vec p}|$ is given by
\cite{collins,tucci,collinssterman,berger,frederix,nayakqed,scet1,pathorder,nayaka2,nayaka3}
\bea
gT^a~\frac{p^\mu}{p \cdot k+i\epsilon }.
\label{eikonalinp}
\eea
In QCD the soft divergences arise only from the
emission of a gluon for which all components of the four-momentum $k^\mu$ are small
($|{\vec k}| \rightarrow 0$) which is evident from eq. (\ref{eikonalinp}). From eq. (\ref{eikonalinp})
one also finds that when $0<|{\vec k}|<<< |{\vec p}|$ and ${\vec p}$ is parallel to ${\vec k}$ one may
find collinear divergences in QCD.

However, since
\bea
p \cdot k =p_0 k_0 -{\vec p} \cdot {\vec k}=|{\vec p}||{\vec k}|(\sqrt{1+\frac{m^2}{{\vec p}^2}}  -{\rm cos}\theta)
\label{scdq}
\eea
one finds that the collinear divergences does not appear when quark interacts with a collinear gluon
because of the non-vanishing mass of the quark, {\it i. e.}, $m\neq 0$ even if the mass of the light quark is
very small. From eq. (\ref{scdq}) one finds that the collinear divergences appear only
when $m=0$ and $\theta =0$ where $\theta$ is the angle between ${\vec p}$ and ${\vec k}$.
Since gluons are massless and the massless gluons interact with each other one finds that the collinear
divergences appear in QCD. Since a massless particle is always light-like one
finds that the soft-collinear divergences can be described by light-like Wilson line in QCD.

For light-like quark we find from eq. (\ref{eikonalinp})
\bea
gT^a~\frac{p^\mu}{p \cdot k+i\epsilon }=gT^a~\frac{l^\mu}{l \cdot k+ i\epsilon }
\label{eikonalin}
\eea
where $l^\mu$ is the light-like four-velocity ($|{\vec l}|$=1) of the quark.
Note that when we say the "light-like quark" we mean the quark
that is traveling at its highest speed which is arbitrarily close to the speed of light
($|{\vec l}|\sim 1$) as it can not travel exactly at speed of light
because it has finite mass even if the mass of the light quark is very small.
On the other hand a massless gluon is light-like and it always remains light-like. Hence we find that
if $l^\mu$ is light-like four velocity then the soft-collinear divergences in QCD can be described by
the Eikonal Feynman rule as given by eq. (\ref{eikonalin}). Note that the Eikonal Feynman rule in eq.
(\ref{eikonalin}) is also valid if we replace the light-like quark by light-like gluon provided we replace
$T^{a}_{bc}=-if^{abc}$.

From eq. (\ref{eikonalin}) we find
\bea
&& gT^a\int \frac{d^4k}{(2\pi)^4} \frac{l\cdot { A}^a(k)}{l\cdot k +i\epsilon } =-gT^a i\int_0^{\infty} d\lambda \int \frac{d^4k}{(2\pi)^4} e^{i l \cdot k \lambda} l\cdot { A}^a(k) = igT^a\int_0^{\infty} d\lambda l\cdot {  A}^a(l\lambda)\nonumber \\
\label{ftgt}
\eea
where the gluon field $ { A}^{\mu a}(x)$ and its Fourier transform $ { A}^{\mu a}(k)$ are related by
\bea
{ A}^{\mu a}(x) =\int \frac{d^4k}{(2\pi)^4} { A}^{\mu a}(k) e^{ik \cdot x}.
\label{ft}
\eea
Note that a path ordering in QCD is required which can be seen as follows, see also \cite{bodwin}. The
Eikonal contribution for the soft-collinear divergences in QCD arising from a single soft-collinear gluon exchange in Feynman diagram
is given by eq. (\ref{ftgt}). Now consider the corresponding Feynman diagram for the soft-collinear divergences in QCD
due to exchange of two soft-collinear gluons of four-momenta $k^\mu_1$ and $k^\mu_2$.
The corresponding Eikonal contribution due to two soft-collinear gluons exchange is analogously given by
\bea
&& g^2\int \frac{d^4k_1}{(2\pi)^4} \frac{d^4k_2}{(2\pi)^4} \frac{T^a l\cdot { A}^a(k_2)T^b l\cdot { A}^b(k_1)}{(l\cdot (k_1+k_2) +i \epsilon)(l\cdot k_1 +i \epsilon)} \nonumber \\
&&=g^2i^2 \int_0^{\infty}  d\lambda_2 \int_{\lambda_2}^{\infty} d\lambda_1 T^a l\cdot { A}^a(l\lambda_2) T^b l\cdot { A}^b(l\lambda_1)
\nonumber \\
&&= \frac{g^2i^2}{2!} {\cal P}\int_0^{\infty}  d\lambda_2 \int_0^{\infty}  d\lambda_1 T^a l\cdot { A}^a(l\lambda_2) T^b l\cdot { A}^b(l\lambda_1)
\eea
where ${\cal P}$ is  the path ordering.
Extending this calculation up to infinite number of soft-collinear gluons we find that the Eikonal contribution for the soft-collinear
divergences due to soft-collinear gluons exchange with the light-like quark in QCD is given by the path ordered exponential
\bea
{\cal P}~{\rm exp}[ig \int_0^{\infty} d\lambda l\cdot { A}^a(l\lambda)T^a ]
\label{iiij}
\eea
where $l^\mu$ is the light-like four velocity of the quark. The Wilson line in QCD is given by
\bea
{\cal P}e^{ig \int_{x_i}^{x_f} dx^\mu A_\mu^a(x)T^a }
\label{tts}
\eea
which is the solution of the equation \cite{lam}
\bea
\partial_\mu S(x)=igT^aA_\mu^a(x)S(x)
\eea
with initial condition
\bea
S(x_i)=1.
\eea
When $A^{\mu a}(x)=A^{\mu a}(l\lambda)$ as in eq. (\ref{iiij}) we find from eq. (\ref{tts}) that the light-like Wilson line in
QCD for soft-collinear divergences is given by \cite{stermanpath}
\bea
{\cal P}e^{ig \int_0^x dx^\mu A_\mu^a(x) T^a}=\left[{\cal P}e^{-ig \int_0^{\infty} d\lambda l\cdot { A}^a(x+l\lambda) T^a}\right]{\cal P}e^{ig \int_0^{\infty} d\lambda l\cdot { A}^a(l\lambda) T^a}.
\label{oh}
\eea

A light-like quark traveling with light-like four-velocity $l^\mu$ produces SU(3) pure gauge potential $A^{\mu a}(x)$
at all the time-space position $x^\mu$ except at the position ${\vec x}$ perpendicular to the direction of motion
of the quark (${\vec l}\cdot {\vec x}=0$) at the time of closest approach \cite{collinssterman,nayakj,nayake}.
When $A^{\mu a}(x) = A^{\mu a}(\lambda l)$ as in eq. (\ref{iiij})
we find ${\vec l}\cdot {\vec x}=\lambda {\vec l}\cdot {\vec l}=\lambda\neq 0$ which implies that the light-like Wilson line
finds the gluon field $A^{\mu a}(x)$ in eq. (\ref{iiij}) as the SU(3) pure gauge. The SU(3) pure gauge is given by
\bea
T^aA_\mu^a (x)= \frac{1}{ig}[\partial_\mu U(x)] ~U^{-1}(x),~~~~~~~~~~~~~U(x)=e^{igT^a\omega^a(x)}
\label{gtqcd}
\eea
which gives
\bea
U(x_f)={\cal P}e^{ig \int_{x_i}^{x_f} dx^\mu A_\mu^a(x) T^a}U(x_i)=e^{igT^a\omega^a(x_f)}.
\label{uxf}
\eea
Hence when $A^{\mu a}(x) = A^{\mu a}(\lambda l)$ as in eq. (\ref{iiij}) we find from eqs. (\ref{oh}) and
(\ref{uxf}) that the light-like Wilson line in QCD for soft-collinear divergences is given by
\bea
{\cal P}e^{ie \int_{0}^{x} dx^\mu A_\mu^a(x)T^a }=e^{igT^a\omega(x)}e^{-igT^a\omega^a(0)}=\left[{\cal P}e^{-ig \int_{0}^{\infty} d\lambda l\cdot { A}^a(x+l\lambda)T^a }\right]{\cal P}e^{ig \int_{0}^{\infty} d\lambda v\cdot { A}^a(l\lambda)T^a }
\label{lkjn}
\eea
which depends only on end points $0$ and $x^\mu$ but is independent of the path.
The path independence can also be found from the non-abelian Stokes theorem which can be seen as follows.
The SU(3) pure gauge in eq. (\ref{gtqcd})
gives
\bea
F^a_{\mu \nu}[A]=\partial_\mu A^a_\nu(x) - \partial_\nu A^a_\mu(x)+gf^{abc} A^b_\mu(x) A^c_\nu(x)=0.
\label{cfmn}
\eea
Using eq. (\ref{cfmn}) in the non-abelian Stokes theorem \cite{stokes} we find
\bea
{\cal P}e^{ig \oint_C dx^\mu A_\mu^a(x)T^a} = {\cal P}{\rm exp}[ig \int_S dx^\mu dx^\nu \left[{\cal P}e^{ig \int_y^x dx'^\lambda A_\lambda^b(x')T^b}\right]F_{\mu \nu}^a(x)  T^a\left[{\cal P}e^{ig \int_x^y dx''^\delta A_\delta^c(x'')T^c}\right]]=1\nonumber \\
\label{555}
\eea
where $C$ is a closed path and $S$ is the surface enclosing $C$. Now considering two different paths
$L$ and $M$ with common end points $0$ and $x^\mu$ we find from eq. (\ref{555})
\bea
&&{\cal P}e^{ig \oint_C dx^\mu A_\mu^a(x)T^a} ={\cal P}{\rm exp}[ig \int_L dx^\mu A_\mu^a(x)T^a- ig \int_M dx^\mu A_\mu^a(x)T^a]\nonumber \\
&&=\left[{\cal P}e^{ig \int_L dx^\mu A_\mu^a(x)T^a}\right]\left[{\cal P}e^{- ig \int_M dx^\nu A_\nu^b(x)T^b}\right]=1
\label{pht}
\eea
which implies that the light-like Wilson line in QCD
\bea
{\cal P}e^{ig \int_0^x dx^\mu A_\mu^a(x)T^a }
\label{llwl}
\eea
depends only on the end points $0$ and $x^\mu$ but is independent of the path which can also be seen from eq. (\ref{lkjn}).
Hence from eq. (\ref{lkjn}) we find that the non-abelian phase or the gauge link in QCD is given by
\bea
\Phi(x)={\cal P}e^{-ig \int_0^{\infty} d\lambda l\cdot { A}^a(x+l\lambda)T^a }=e^{igT^a\omega^a(x)}.
\label{ttt}
\eea

Note that from eq. (\ref{cfmn}) we find the vanishing physical gauge invariant field strength square $F^{\mu \nu a}[A]F^a_{\mu \nu}[A]$
when $A^{\mu a}(x)$ is the SU(3) pure gauge as given by eq. (\ref{gtqcd}).
Hence in classical mechanics the SU(3) pure gauge potential does not have an effect on color charged
particle and one expects the effect of exchange of soft-collinear gluons to simply vanish.
However, in quantum mechanics the situation is a little more complicated, because the gauge potential
does have an effect on color charged particle even if it is SU(3) pure gauge potential and hence
one should not expect the effect of exchange of soft-collinear gluons to simply vanish \cite{collinssterman}.
This can be verified by studying the non-perturbative correlation function
of the type $<0|{\bar \psi}(x) \psi(x') {\bar \psi}(x'') \psi(x''')...|0>$ in QCD in the
presence of SU(3) pure gauge background field.

Under non-abelian gauge transformation given by eq. (\ref{aftgrmpi})
the Wilson line in QCD transforms as
\bea
{\cal P}e^{ie \int_{x_i}^{x_f} dx^\mu A'^a_\mu(x)T^a }=U(x_f)\left[{\cal P}e^{ie \int_{x_i}^{x_f} dx^\mu A^a_\mu(x)T^a }\right]U^{-1}(x_i).
\label{ty}
\eea
From eqs. (\ref{lkjn}) and (\ref{ty}) we find
\bea
{\cal P}e^{-ig \int_0^{\infty} d\lambda l\cdot { A}'^a(x+l\lambda)T^a }=U(x){\cal P}e^{-ig \int_0^{\infty} d\lambda l\cdot { A}^a(x+l\lambda)T^a },~~~~~~~~~~~~~~U(x)={\rm exp}[igT^a\omega^a(x)]\nonumber \\
\label{kkkj}
\eea
which gives from eq. (\ref{ttt})
\bea
\Phi'(x)=U(x)\Phi(x),~~~~~~~~~~~~~~~~~~~~~\Phi'^\dagger(x)=\Phi^\dagger(x)U^{-1}(x).
\label{hqg}
\eea

In the adjoint representation of SU(3) the corresponding path ordered exponential is given by
\bea
\Phi_{ab}(x)={\cal P}{\rm exp}[-ig\int_0^{\infty} d\lambda l\cdot { A}^c(x+l\lambda)T^{(A)c}]=e^{igT^{(A)c}\omega^c(x)},~~~~~~(T^{(A)c})_{ab}=-if^{abc}.\nonumber \\
\label{adjin}
\eea

To summarize this, we find that the soft-collinear divergences in the perturbative Feynman diagrams due to soft-collinear gluons interaction
with the light-like Wilson line in QCD is given by the path ordered exponential in eq. (\ref{iiij}) which is nothing
but the non-abelian phase or the gauge link in QCD as given by eq. (\ref{ttt}) where the gluon field $A^{\mu a}(x)$ is the
SU(3) pure gauge, see eqs. (\ref{gtqcd}), (\ref{uxf}), (\ref{lkjn}). This implies that the effect of soft-collinear gluons
interaction between the partons and the light-like Wilson line in QCD can be studied by
putting the partons in the SU(3) pure gauge background field.
Hence we find that the soft-collinear behavior of the non-perturbative correlation function
of the type $<0|{\bar \psi}(x) \psi(x') {\bar \psi}(x'') \psi(x''')...|0>$ in QCD
due to the presence of light-like Wilson line in QCD can be studied by using the path integral method
of the QCD in the presence of SU(3) pure gauge background field.

It can be mentioned here that in soft collinear effective theory
(SCET) \cite{scet} it is also necessary to use the idea of background fields \cite{abbott} to give well defined meaning to several
distinct gluon fields \cite{scet1}.

Note that a massive color source traveling at speed much less than speed of light
can not produce SU(3) pure gauge field \cite{collinssterman,nayakj,nayake}. Hence when one replaces light-like
Wilson line with massive Wilson line one expects the factorization of soft/infrared divergences to
break down. This is in conformation with the finding in \cite{nayaksterman1} which used the diagrammatic
method of QCD. In case of massive Wilson line in QCD the color transfer occurs and the factorization breaks
down. Note that in case of massive Wilson line there is no collinear divergences which is explained in eq.
(\ref{scdq}).

\section{ Non-perturbative gluon correlation function and proof of factorization theorem in covariant gauge }

Since gluon distribution function inside the hadron is a non-perturbative quantity in QCD
it is natural to use path integral method of QCD to study its properties from first principles.
The generating functional in the path integral method of QCD is given by \cite{muta,abbott}
\bea
Z[J,\eta,{\bar \eta}]=\int [dQ] [d{\bar \psi}] [d \psi ] ~{\rm det}(\frac{\delta (\partial_\mu Q^{\mu a})}{\delta \omega^b})
~e^{i\int d^4x [-\frac{1}{4}{F^a}_{\mu \nu}^2[Q] -\frac{1}{2 \alpha} (\partial_\mu Q^{\mu a})^2+{\bar \psi} [i\gamma^\mu \partial_\mu -m +gT^a\gamma^\mu Q^a_\mu] \psi + J \cdot Q +{\bar \eta} \psi +  {\bar \psi} \eta]} \nonumber \\
\label{zfq}
\eea
where $J^{\mu a}(x)$ is the external source for the quantum gluon field $Q^{\mu a}(x)$ and ${\bar \eta}_i(x)$ is the external source for the
Dirac field $\psi_i(x)$ of the quark and
\bea
F^a_{\mu \nu}[Q]=\partial_\mu Q^a_\nu(x)-\partial_\nu Q^a_\mu(x)+gf^{abc}Q^b_\mu(x)Q^c_\nu(x),~~~~~~~~~{F^a}_{\mu \nu}^2[Q]={F}^{\mu \nu a}[Q]{F}^a_{\mu \nu}[Q].
\eea
Note that the Faddeev-Popov (F-P) determinant ${\rm det}(\frac{\delta (\partial_\mu Q^{\mu a})}{\delta \omega^b})$ can be expressed in terms of path
integral over the ghost fields \cite{muta} but we will directly work with the Faddeev-Popov (F-P) determinant
${\rm det}(\frac{\delta (\partial_\mu Q^{\mu a})}{\delta \omega^b})$ in this paper.
The non-perturbative correlation function of the type $<0|{\bar \psi}(x_1) \psi(x_2)|0>$ in QCD is given by \cite{tucci}
\bea
&&<0|{\bar \psi}(x_1) \psi(x_2)|0>=\frac{1}{Z[0]}\int [dQ] [d{\bar \psi}] [d \psi ] ~{\bar \psi}(x_1) \psi(x_2)\nonumber \\
&& \times {\rm det}(\frac{\delta (\partial_\mu Q^{\mu a})}{\delta \omega^b})~
e^{i\int d^4x [-\frac{1}{4}{F^a}_{\mu \nu}^2[Q] -\frac{1}{2 \alpha}(\partial_\mu Q^{\mu a})^2+{\bar \psi} [i\gamma^\mu \partial_\mu -m +gT^a\gamma^\mu Q^a_\mu] \psi  ]}.
\label{corq}
\eea
Similarly the non-perturbative gluon correlation function of the type $<0|Q_\mu^a(x_1) Q_\nu^b(x_2)|0>$ in QCD is given by \cite{tucci}
\bea
&&<0|Q_\mu^a(x_1) Q_\nu^b(x_2)|0>=\int [dQ] [d{\bar \psi}] [d \psi ] ~Q_\mu^a(x_1) Q_\nu^b(x_2)\nonumber \\
&& \times {\rm det}(\frac{\delta (\partial_\mu Q^{\mu a})}{\delta \omega^b})~
e^{i\int d^4x [-\frac{1}{4}{F^a}_{\mu \nu}^2[Q] -\frac{1}{2 \alpha}(\partial_\mu Q^{\mu a})^2+{\bar \psi} [i\gamma^\mu \partial_\mu -m +gT^a\gamma^\mu Q^a_\mu] \psi  ]}
\label{cfq5}
\eea
where the suppression of the normalization factor $Z[0]$ is understood as it will cancel in the final result (see for 
example, eq. (\ref{finaly})). 

We have seen in section II that the soft-collinear behavior of the non-perturbative (gluon) correlation function
due to the presence of light-like Wilson line in QCD can be studied by using the path integral method
of the QCD in the presence of SU(3) pure gauge background field. Hence in order to derive correct definition of the
gluon distribution function at high energy colliders we use the
path integral formulation of the background field method of QCD
in the presence of SU(3) pure gauge background field as given by eq. (\ref{gtqcd}).

Background field method of QCD was originally formulated by 't Hooft \cite{thooft} and later
extended by Klueberg-Stern and Zuber \cite{zuber,zuber1} and by Abbott \cite{abbott}.
This is an elegant formalism which can be useful to construct gauge invariant
non-perturbative green's functions in QCD. This formalism is also useful to study quark and gluon production from classical chromo field \cite{peter}
via Schwinger mechanism \cite{schw}, to compute $\beta$ function in QCD \cite{peskin}, to perform
calculations in lattice gauge theories \cite{lattice} and to study evolution of QCD
coupling constant in the presence of chromofield \cite{nayak}.

The generating functional in the path integral formulation of the background field method of QCD is
given by \cite{thooft,abbott,zuber}
\bea
&& Z[A,J,\eta,{\bar \eta}]=\int [dQ] [d{\bar \psi}] [d \psi ] ~{\rm det}(\frac{\delta G^a(Q)}{\delta \omega^b}) \nonumber \\
&& e^{i\int d^4x [-\frac{1}{4}{F^a}_{\mu \nu}^2[A+Q] -\frac{1}{2 \alpha}
(G^a(Q))^2+{\bar \psi} [i\gamma^\mu \partial_\mu -m +gT^a\gamma^\mu (A+Q)^a_\mu] \psi + J \cdot Q +{\bar \eta} \psi + {\bar \psi}\eta ]}
\label{azaqcd}
\eea
where the gauge fixing term is given by
\bea
G^a(Q) =\partial_\mu Q^{\mu a} + gf^{abc} A_\mu^b Q^{\mu c}=D_\mu[A]Q^{\mu a}
\label{ga}
\eea
which depends on the background field $A^{\mu a}(x)$ and
\bea
F_{\mu \nu}^a[A+Q]=\partial_\mu [A_\nu^a+Q_\nu^a]-\partial_\nu [A_\mu^a+Q_\mu^a]+gf^{abc} [A_\mu^b+Q_\mu^b][A_\nu^c+Q_\nu^c].
\eea
We have followed the notations of \cite{thooft,zuber,abbott} and accordingly we have
denoted the quantum gluon field by $Q^{\mu a}$ and the background field by $A^{\mu a}$.
Note that in the absence of the external sources the SU(3) pure gauge can be gauged away from the generating functional in the background
field method of QCD. However, in the presence of the external sources the SU(3) pure gauge can not be gauged away from the generating
functional in the background field method of QCD.

The non-perturbative correlation function
of the type $<0|{\bar \psi}(x_1) \psi(x_2)|0>_A$ in the background field method of QCD is given by \cite{tucci}
\bea
&&<0|{\bar \psi}(x_1) \psi(x_2)|0>_A=\frac{1}{Z[0]}\int [dQ] [d{\bar \psi}] [d \psi ] ~{\bar \psi}(x_1) \psi(x_2)\nonumber \\
&& \times {\rm det}(\frac{\delta G^a(Q)}{\delta \omega^b}) e^{i\int d^4x [-\frac{1}{4}{F^a}_{\mu \nu}^2[A+Q] -\frac{1}{2 \alpha}
(G^a(Q))^2+{\bar \psi} [i\gamma^\mu \partial_\mu -m +gT^a\gamma^\mu (A+Q)^a_\mu] \psi   ]}.
\label{corqa}
\eea
Similarly the non-perturbative gluon correlation function of the type $<0|Q_\mu^a(x_1) Q_\nu^b(x_2)|0>_A$
in the background field method of QCD is given by \cite{tucci}
\bea
&&<0|Q_\mu^a(x_1) Q_\nu^b(x_2)|0>_A=\int [dQ] [d{\bar \psi}] [d \psi ] ~Q_\mu^a(x_1) Q_\nu^b(x_2)\nonumber \\
&& \times {\rm det}(\frac{\delta G^a(Q)}{\delta \omega^b}) e^{i\int d^4x [-\frac{1}{4}{F^a}_{\mu \nu}^2[A+Q] -\frac{1}{2 \alpha}
(G^a(Q))^2+{\bar \psi} [i\gamma^\mu \partial_\mu -m +gT^a\gamma^\mu (A+Q)^a_\mu] \psi   ]}
\label{cfqcd}
\eea
where the suppression of the normalization factor $Z[0]$ is understood as it will cancel in the final result (see for
example, eq. (\ref{finaly})). 

The gauge fixing term $\frac{1}{2 \alpha} (G^a(Q))^2$ in eq. (\ref{azaqcd}) [where $G^a(Q)$ is given by eq. (\ref{ga})]
is invariant for gauge transformation of $A_\mu^a$:
\bea
\delta A_\mu^a = gf^{abc}A_\mu^b\omega^c + \partial_\mu \omega^a,  ~~~~~~~({\rm type~ I ~transformation})
\label{typeI}
\eea
provided one also performs a homogeneous transformation of $Q_\mu^a$ \cite{zuber,abbott}:
\bea
\delta Q_\mu^a =gf^{abc}Q_\mu^b\omega^c.
\label{omega}
\eea
The gauge transformation of background field $A_\mu^a$ as given by eq. (\ref{typeI})
along with the homogeneous transformation of $Q_\mu^a$ in eq. (\ref{omega}) gives
\bea
\delta (A_\mu^a+Q_\mu^a) = gf^{abc}(A_\mu^b+Q_\mu^b)\omega^c + \partial_\mu \omega^a
\label{omegavbxn}
\eea
which leaves $-\frac{1}{4}{F^a}_{\mu \nu}^2[A+Q]$ invariant in eq. (\ref{azaqcd}).

For fixed $A_\mu^a$, {\it i.e.}, for
\bea
&&\delta A_\mu^a =0,  ~~~~~~~({\rm type~ II ~transformation})
\label{typeII}
\eea
the gauge transformation of $Q_\mu^a$ \cite{zuber,abbott}:
\bea
&&\delta Q_\mu^a = gf^{abc}(A_\mu^b + Q_\mu^b)\omega^c + \partial_\mu \omega^a
\label{omegaII}
\eea
gives eq. (\ref{omegavbxn}) which leaves $-\frac{1}{4}{F^a}_{\mu \nu}^2[A+Q]$ invariant in eq. (\ref{azaqcd}).
Hence whether we use type I transformation [eqs. (\ref{typeI}) and (\ref{omega})] or type II transformation
[eqs. (\ref{typeII}) and (\ref{omegaII})] we will obtain the same equation (\ref{cfq5p2}).

It is useful to remember that, unlike QED \cite{tucci}, finding an exact relation between the generating
functional $Z[J, \eta, {\bar \eta}]$ in QCD in eq. (\ref{zfq}) and the generating functional $Z[A, J, \eta, {\bar \eta}]$
in the background field method of QCD in eq. (\ref{azaqcd}) in the presence of SU(3) pure gauge background field is not easy.
The main difficulty is due to the gauge fixing terms which are different in both the cases. While the Lorentz (covariant) gauge fixing
term  $-\frac{1}{2 \alpha}(\partial_\mu Q^{\mu a})^2$ in eq. (\ref{zfq}) in QCD is independent of the background field
$A^{\mu a}(x)$, the background field gauge fixing term $-\frac{1}{2 \alpha}(G^a(Q))^2$ in eq. (\ref{azaqcd}) in the background field method
of QCD depends on the background field $A^{\mu a}(x)$ where $G^a(Q)$ is given by eq. (\ref{ga}) \cite{thooft,zuber,abbott}.
Hence in order to study non-perturbative correlation function of the gluon
in the background field method of QCD in the presence of SU(3) pure gauge background
field we proceed as follows.

By changing the integration variable $Q \rightarrow Q-A$ in the right hand side of eq. (\ref{cfqcd}) we find
\bea
&& <0|Q_\mu^a(x_1) Q_\nu^b(x_2)|0>_A= \int [dQ] [d{\bar \psi}] [d \psi ]  ~[Q_\mu^a(x_1)- A_\mu^a(x_1) ] [ Q_\nu^b(x_2)- A_\nu^b(x_2)] \nonumber \\
&& \times {\rm det}(\frac{\delta G_f^a(Q)}{\delta \omega^b})~~
e^{i\int d^4x [-\frac{1}{4}{F^a}_{\mu \nu}^2[Q] -\frac{1}{2 \alpha} (G_f^a(Q))^2+{\bar \psi} [i\gamma^\mu \partial_\mu -m +gT^a\gamma^\mu Q^a_\mu] \psi  ]}
\label{cfqcd1}
\eea
where from eq. (\ref{ga}) we find
\bea
G_f^a(Q) =\partial_\mu Q^{\mu a} + gf^{abc} A_\mu^b Q^{\mu c} - \partial_\mu A^{\mu a}=D_\mu[A] Q^{\mu a} - \partial_\mu A^{\mu a}
\label{gfa}
\eea
and from eq. (\ref{omega}) [by using eq. (\ref{typeI})] we find
\bea
\delta Q_\mu^a = -gf^{abc}\omega^b Q_\mu^c + \partial_\mu \omega^a.
\label{theta}
\eea
The eqs. (\ref{cfqcd1}), (\ref{gfa}) and (\ref{theta}) can also be derived by using type II transformation
which can be seen as follows. By changing $Q \rightarrow Q-A$ in eq. (\ref{cfqcd}) we find eq. (\ref{cfqcd1})
where the gauge fixing term from eq. (\ref{ga}) becomes eq. (\ref{gfa})
and eq. (\ref{omegaII}) [by using eq. (\ref{typeII})] becomes eq. (\ref{theta}).
Hence we obtain eqs. (\ref{cfqcd1}), (\ref{gfa}) and (\ref{theta}) whether we use
the type I transformation or type II transformation. Hence we find that we will obtain the same
eq. (\ref{cfq5p2}) whether we use the type I transformation or type II transformation.

Changing the integration variable from unprimed variable to primed variable we find from eq. (\ref{cfqcd1})
\bea
&& <0|Q_\mu^a(x_1) Q_\nu^b(x_2)|0>_A= \int [dQ'] [d{\bar \psi}'] [d \psi' ]~[Q_\mu'^a(x_1)- A_\mu^a(x_1) ] [ Q_\nu'^b(x_2)- A_\nu^b(x_2)] \nonumber \\
&& \times {\rm det}(\frac{\delta G_f^a(Q')}{\delta \omega^b})~~
e^{i\int d^4x [-\frac{1}{4}{F^a}_{\mu \nu}^2[Q'] -\frac{1}{2 \alpha} (G_f^a(Q'))^2+{\bar \psi}' [i\gamma^\mu \partial_\mu -m +gT^a\gamma^\mu Q'^a_\mu] \psi'  ]}.
\label{cfqcd1vb}
\eea
This is because a change of integration variable from unprimed variable to primed variable does not change the value of the
integration.

The equation
\bea
Q'^a_\mu(x)= Q^a_\mu(x) +gf^{abc}\omega^c(x) Q_\mu^b(x) + \partial_\mu \omega^a(x)
\label{thetaav}
\eea
in eq. (\ref{theta}) is valid for infinitesimal transformation ($\omega << 1$) which is obtained from the
finite equation
\bea
T^aQ'^a_\mu(x) = U(x)T^aQ^a_\mu(x) U^{-1}(x)+\frac{1}{ig}[\partial_\mu U(x)] U^{-1}(x),~~~~~~~~~~~U(x)=e^{igT^a\omega^a(x)}.
\label{ftgrm}
\eea
Simplifying infinite numbers of non-commuting terms we find
\bea
\left[~e^{-igT^b\omega^b(x)} ~T^a~ e^{igT^c \omega^c(x)}~\right]_{ij}=[e^{-gM(x)}]_{ab}T^b_{ij}
\label{non}
\eea
where
\bea
M_{ab}(x)=f^{abc}\omega^c(x).
\label{mab}
\eea
Hence by simplifying infinite numbers of non-commuting terms in eq. (\ref{ftgrm}) [by using eq. (\ref{non}) and \cite{nayakj}] we find that
\bea
{Q'}_\mu^a(x) = [e^{gM(x)}]_{ab}Q_\mu^b(x) ~+ ~[\frac{e^{gM(x)}-1}{gM(x)}]_{ab}~[\partial_\mu \omega^b(x)],~~~~~~~~~~~M_{ab}(x)=f^{abc}\omega^c(x).
\label{teq}
\eea

Under the finite transformation, using eq. (\ref{teq}), we find
\bea
&& [dQ'] =[dQ] ~{\rm det} [\frac{\partial {Q'}^a}{\partial Q^b}] = [dQ] ~{\rm det} [[e^{gM(x)}]]=[dQ] {\rm exp}[{\rm Tr}({\rm ln}[e^{gM(x)}])]=[dQ]
\label{dqa}
\eea
where we have used (for any matrix $H$)
\bea
{\rm det}H={\rm exp}[{\rm Tr}({\rm ln}H)].
\eea

Similarly the fermion fields transform accordingly, see eq. (\ref{phg3}).
Using eqs. (\ref{teq}) and (\ref{phg3}) we find
\bea
&&[d{\bar \psi}'] [d \psi' ]=[d{\bar \psi}] [d \psi ],~~~~~~{\bar \psi}' [i\gamma^\mu \partial_\mu -m +gT^a\gamma^\mu Q'^a_\mu] \psi'={\bar \psi} [i\gamma^\mu \partial_\mu -m +gT^a\gamma^\mu Q^a_\mu]\psi,\nonumber \\
&&{F^a}_{\mu \nu}^2[Q']={F^a}_{\mu \nu}^2[Q].
\label{psa}
\eea
Simplifying all the infinite number of non-commuting terms in eq. (\ref{gtqcd}) we find that the
SU(3) pure gauge $A^{\mu a}(x)$ is given by \cite{nayakj}
\bea
A^{\mu a}(x)=\partial^\mu \omega^b(x)\left[\frac{e^{gM(x)}-1}{gM(x)}\right]_{ab}
\label{pg4}
\eea
where $M_{ab}(x)$ is given by eq. (\ref{mab}). From eqs. (\ref{pg4}) and (\ref{teq}) we find
\bea
{Q'}_\mu^a(x) -A_\mu^a(x)= [e^{gM(x)}]_{ab}Q_\mu^b(x),~~~~~~~~~~~M_{ab}(x)=f^{abc}\omega^c(x).
\label{tej}
\eea

Using eqs. (\ref{dqa}), (\ref{psa}) and (\ref{tej}) in eq. (\ref{cfqcd1vb}) we find
\bea
&& <0|Q_\mu^a(x_1) Q_\nu^b(x_2)|0>_A=~\int [dQ] [d{\bar \psi}] [d \psi ] ~[e^{gM(x_1)}]_{ac}Q_\mu^c(x_1)~[e^{gM(x_2)}]_{bd}Q_\nu^d(x_2)\nonumber \\
 && \times {\rm det}(\frac{\delta G_f^a(Q')}{\delta \omega^b})
~e^{i\int d^4x [-\frac{1}{4}{F^a}_{\mu \nu}^2[Q] -\frac{1}{2 \alpha} (G_f^a(Q'))^2+{\bar \psi} [i\gamma^\mu \partial_\mu -m +gT^a\gamma^\mu Q^a_\mu] \psi ]}.
\label{cfqcd1c}
\eea

From eq. (\ref{gfa}) we find
\bea
G_f^a(Q') =\partial_\mu Q^{' \mu a} + gf^{abc} A_\mu^b Q^{' \mu c} - \partial_\mu A^{\mu a}.
\label{gfap}
\eea
By using eqs. (\ref{teq}) and (\ref{pg4}) in eq. (\ref{gfap}) we find
\bea
&&G_f^a(Q') =\partial^\mu [[e^{gM(x)}]_{ab}Q_\mu^b(x) ~+ ~[\frac{e^{gM(x)}-1}{gM(x)}]_{ab}~[\partial_\mu \omega^b(x)]]\nonumber \\
&&+ gf^{abc} [\partial^\mu \omega^e(x)\left[\frac{e^{gM(x)}-1}{gM(x)}\right]_{be}] [[e^{gM(x)}]_{cd}Q_\mu^d(x) ~+ ~[\frac{e^{gM(x)}-1}{gM(x)}]_{cd}~[\partial_\mu \omega^d(x)]]\nonumber \\
&&- \partial_\mu [\partial^\mu \omega^b(x)\left[\frac{e^{gM(x)}-1}{gM(x)}\right]_{ab}]
\label{gfapa}
\eea
which gives
\bea
&&G_f^a(Q') =\partial^\mu [[e^{gM(x)}]_{ab}Q_\mu^b(x)]\nonumber \\
&&+ gf^{abc} [\partial^\mu \omega^e(x)\left[\frac{e^{gM(x)}-1}{gM(x)}\right]_{be}] [[e^{gM(x)}]_{cd}Q_\mu^d(x) ~+ ~[\frac{e^{gM(x)}-1}{gM(x)}]_{cd}~[\partial_\mu \omega^d(x)]].
\label{gfapb}
\eea
From eq. (\ref{gfapb}) we find
\bea
&&G_f^a(Q') =\partial^\mu [[e^{gM(x)}]_{ab}Q_\mu^b(x)]+ gf^{abc} [[\partial^\mu \omega^e(x)]\left[\frac{e^{gM(x)}-1}{gM(x)}\right]_{be}] [[e^{gM(x)}]_{cd}Q_\mu^d(x)]
\label{gfapc}
\eea
which gives
\bea
&&G_f^a(Q') = [e^{gM(x)}]_{ab}\partial^\mu Q_\mu^b(x)\nonumber \\
&&+Q_\mu^b(x)\partial^\mu [[e^{gM(x)}]_{ab}]+  [[\partial^\mu \omega^e(x)]\left[\frac{e^{gM(x)}-1}{gM(x)}\right]_{be}]gf^{abc} [[e^{gM(x)}]_{cd}Q_\mu^d(x)].
\label{gfapd}
\eea
From \cite{nayakj} we find
\bea
\partial^\mu [e^{igT^a\omega^a(x)}]_{ij}=ig[\partial^\mu \omega^b(x)]\left[\frac{e^{gM(x)}-1}{gM(x)}\right]_{ab}T^a_{ik}[e^{igT^c\omega^c(x)}]_{kj},~~~~~~~~~M_{ab}(x)=f^{abc}\omega^c(x)\nonumber \\
\label{pg4j}
\eea
which in the adjoint representation of SU(3) gives (by using $T^a_{bc}=-if^{abc}$)
\bea
[\partial^\mu e^{gM(x)}]_{ad}=[\partial^\mu \omega^e(x)]\left[\frac{e^{gM(x)}-1}{gM(x)}\right]_{be}gf^{bac}[e^{M(x)}]_{cd},~~~~~~~~~M_{ab}(x)=f^{abc}\omega^c(x).
\label{pg4k}
\eea
Using eq. (\ref{pg4k}) in (\ref{gfapd}) we find
\bea
&&G_f^a(Q') = [e^{gM(x)}]_{ab}\partial^\mu Q_\mu^b(x)
\label{gfape}
\eea
which gives
\bea
(G_f^a(Q'))^2 = (\partial_\mu Q^{\mu a}(x))^2.
\label{gfapf}
\eea
Since for $n \times n$ matrices $A$ and $B$ we have
\bea
{\rm det}(AB)=({\rm det}A)({\rm det} B)
\label{detr}
\eea
we find by using eq. (\ref{gfape}) that
\bea
&&{\rm det} [\frac{\delta G_f^a(Q')}{\delta \omega^b}] ={\rm det}
[\frac{ \delta [[e^{gM(x)}]_{ac}\partial^\mu Q_\mu^c(x)]}{\delta \omega^b}]={\rm det}[
[e^{gM(x)}]_{ac}\frac{ \delta (\partial^\mu Q_\mu^c(x))}{\delta \omega^b}]\nonumber \\
&&=\left[{\rm det}[
[e^{gM(x)}]_{ac}]\right]~\left[{\rm det}[\frac{ \delta (\partial^\mu Q_\mu^c(x))}{\delta \omega^b}]\right]={\rm exp}[{\rm Tr}({\rm ln}[e^{gM(x)}])]~{\rm det}[\frac{ \delta (\partial_\mu Q^{\mu a}(x))}{\delta \omega^b}]\nonumber \\
&&={\rm det}[\frac{ \delta (\partial_\mu Q^{\mu a}(x))}{\delta \omega^b}].
\label{gqp4a}
\eea
Using eqs. (\ref{gfapf}) and (\ref{gqp4a}) in eq. (\ref{cfqcd1c}) we find
\bea
&&<0|Q_\mu^a(x_1) Q_\nu^b(x_2)|0>_A=\int [dQ] [d{\bar \psi}] [d \psi ] ~[e^{gM(x_1)}]_{ac}Q_\mu^c(x_1)~[e^{gM(x_2)}]_{bd}Q_\nu^d(x_2)\nonumber \\
 && \times
{\rm det}(\frac{\delta (\partial_\mu Q^{\mu a})}{\delta \omega^b})~
e^{i\int d^4x [-\frac{1}{4}{F^a}_{\mu \nu}^2[Q] -\frac{1}{2 \alpha}(\partial_\mu Q^{\mu a})^2+{\bar \psi} [i\gamma^\mu \partial_\mu -m +gT^a\gamma^\mu Q^a_\mu] \psi  ]}.
\label{cfq5p1}
\eea
Using the similar technique as above we find
\bea
&&
<0|[e^{gM(x_1)}]_{ac}Q_\mu^c(x_1) [e^{gM(x_2)}]_{bd}Q_\nu^d(x_2)|0>_A
 =\int [dQ] [d{\bar \psi}] [d \psi ] ~Q_\mu^a(x_1) Q_\nu^b(x_2)\nonumber \\
 && \times {\rm det}(\frac{\delta (\partial_\mu Q^{\mu a})}{\delta \omega^b})~
e^{i\int d^4x [-\frac{1}{4}{F^a}_{\mu \nu}^2[Q] -\frac{1}{2 \alpha}(\partial_\mu Q^{\mu a})^2+{\bar \psi} [i\gamma^\mu \partial_\mu -m +gT^a\gamma^\mu Q^a_\mu] \psi  ]}
\label{cfq5p2}
\eea
in the presence of SU(3) pure gauge background field $A^{\mu a}(x)$ as given by eq. (\ref{gtqcd}) where $M_{ab}(x)$
is given by eq. (\ref{mab}).

Note that eq. (\ref{cfq5p2}) is valid whether we use type I transformation [eqs. (\ref{typeI}) and (\ref{omega})] or type II transformation
[eqs. (\ref{typeII}) and (\ref{omegaII})]. However, since eq. (\ref{aftgrmpi}) is used to study the gauge transformation of the Wilson line in
QCD as given by eq. (\ref{ty}), we will use type I transformation [see eqs. (\ref{typeI}) and (\ref{omega})] in the rest of the paper
which for the finite transformation give eq. (\ref{aftgrmpi}) and (\ref{jprtpi}) \cite{abbott,zuber}. From eq.
(\ref{jprtpi}) we find that under gauge transformation given by eq. (\ref{aftgrmpi}) the (quantum) gluon field $Q^{\mu a}(x)$ transforms as
\bea
Q'^a_\mu(x)=[e^{gM(x)}]_{ab}Q_\mu^b(x)
\label{jpr}
\eea
where $M_{ab}(x)$ is given by eq. (\ref{mab}). From eqs. (\ref{cfq5}) and (\ref{cfq5p2}) we find
\bea
<0|Q_\mu^a(x_1)Q_\nu^b(x_2)|0> =<0|[e^{gM(x_1)}]_{ac}Q_\mu^c(x_1) [e^{gM(x_2)}]_{bd}Q_\nu^d(x_2)|0>_A.
\label{finaly}
\eea
From eqs. (\ref{mab}), (\ref{adjin}) and (\ref{finaly}) we find
\bea
<0|Q_\mu^a(x_1)Q_\nu^b(x_2)|0> =<0|\Phi_{ac}(x_1)Q_\mu^c(x_1) \Phi_{bd}(x_2)Q_\nu^d(x_2)|0>_A
\label{finalz}
\eea
which proves factorization of soft-collinear divergences at all order in coupling constant in QCD where
\bea
\Phi_{ab}(x)={\cal P}{\rm exp}[-ig\int_0^{\infty} d\lambda l\cdot { A}^c(x+l\lambda)T^{(A)c}],~~~~~~(T^{(A)c})_{ab}=-if^{abc}
\label{wilabf}
\eea
is the non-abelian gauge link or non-abelian phase in the adjoint representation of SU(3) and
$l^\mu$ is the light-like four velocity.

\section{ Non-perturbative gluon correlation function and proof of factorization theorem in general axial gauges }

In QCD the generating functional with general axial gauge fixing is given by \cite{leib,meis}
\bea
&& Z[J,\eta,{\bar \eta}]=\int [dQ] [d{\bar \psi}] [d \psi ] \nonumber \\
&& \times e^{i\int d^4x [-\frac{1}{4}{F^a}_{\mu \nu}^2[Q] -\frac{1}{2 \alpha} (\eta_\mu Q^{\mu a})^2+{\bar \psi} [i\gamma^\mu \partial_\mu -m +gT^a\gamma^\mu Q^a_\mu]  \psi + J \cdot Q +{\bar \psi} \eta +{\bar \eta} \psi ]}
\label{zfqa}
\eea
where $\eta^\mu$ is an arbitrary but constant four vector
\bea
&&\eta^2 <0,~~~~~~~~{\rm axial~gauge}\nonumber \\
&& \eta^2 =0,~~~~~~~~{\rm light}-{\rm cone~gauge}\nonumber \\
&& \eta^2 >0,~~~~~~~~{\rm temporal~gauge}.
\label{axial}
\eea
Note that unlike covariant gauge in eq. (\ref{zfq}) there is no Faddeev-Popov (F-P) determinant in eq. (\ref{zfqa}) because the
ghost particles decouple in general axial gauges \cite{leib,meis}.
The non-perturbative gluon correlation function of the type $<0|Q_\mu^a(x_1) Q_\nu^b(x_2)|0>$ in QCD in general axial gauges
is given by
\bea
&&<0|Q_\mu^a(x_1) Q_\nu^b(x_2)|0>=\int [dQ] [d{\bar \psi}] [d \psi ] ~Q_\mu^a(x_1) Q_\nu^b(x_2)\nonumber \\
&& \times
e^{i\int d^4x [-\frac{1}{4}{F^a}_{\mu \nu}^2[Q] -\frac{1}{2 \alpha} (\eta^\mu Q_\mu^a)^2+{\bar \psi} [i\gamma^\mu \partial_\mu -m +gT^a\gamma^\mu Q^a_\mu] \psi  ]}.
\label{cfq5a}
\eea

The generating functional in the background field method of QCD with general axial gauge fixing is given by \cite{meis}
\bea
&& Z[A,J,\eta,{\bar \eta}]=\int [dQ] [d{\bar \psi}] [d \psi ] \nonumber \\
&& \times e^{i\int d^4x [-\frac{1}{4}{F^a}_{\mu \nu}^2[A+Q] -\frac{1}{2 \alpha}
(\eta^\mu Q_\mu^a)^2+{\bar \psi} [i\gamma^\mu \partial_\mu -m +gT^a\gamma^\mu (A+Q)^a_\mu] \psi + J \cdot Q +{\bar \eta} \psi +{\bar \psi} \eta  ]}.
\label{azaqcda}
\eea

The non-perturbative gluon correlation function of the type $<0|Q_\mu^a(x_1) Q_\nu^b(x_2)|0>_A$
in the background field method of QCD in general axial gauges is given by
\bea
&&<0|Q_\mu^a(x_1) Q_\nu^b(x_2)|0>_A=\int [dQ] [d{\bar \psi}] [d \psi ] ~Q_\mu^a(x_1) Q_\nu^b(x_2)\nonumber \\
&& \times e^{i\int d^4x [-\frac{1}{4}{F^a}_{\mu \nu}^2[A+Q] -\frac{1}{2 \alpha}
(\eta^\mu Q_\mu^a)^2+{\bar \psi} [i\gamma^\mu \partial_\mu -m +gT^a\gamma^\mu (A+Q)^a_\mu] \psi   ]}.
\label{cfqcda}
\eea

By changing the integration variable $Q \rightarrow Q-A$ in the right hand side of eq. (\ref{cfqcda}) we find
\bea
&& <0|Q_\mu^a(x_1) Q_\nu^b(x_2)|0>_A= \int [dQ] [d{\bar \psi}] [d \psi ]  ~[Q_\mu^a(x_1)- A_\mu^a(x_1) ] [ Q_\nu^b(x_2)- A_\nu^b(x_2)] \nonumber \\
&& \times
e^{i\int d^4x [-\frac{1}{4}{F^a}_{\mu \nu}^2[Q] -\frac{1}{2 \alpha} (\eta^\mu (Q-A)_\mu^a)^2+{\bar \psi} [i\gamma^\mu \partial_\mu -m +gT^a\gamma^\mu Q^a_\mu] \psi  ]}.
\label{cfqcd1a}
\eea

Changing the integration variable from unprimed variable to primed variable we find from eq. (\ref{cfqcd1a})
\bea
&& <0|Q_\mu^a(x_1) Q_\nu^b(x_2)|0>_A= \int [dQ'] [d{\bar \psi}'] [d \psi' ]~[Q_\mu'^a(x_1)- A_\mu^a(x_1) ] [ Q_\nu'^b(x_2)- A_\nu^b(x_2)] \nonumber \\
&& \times
e^{i\int d^4x [-\frac{1}{4}{F^a}_{\mu \nu}^2[Q'] -\frac{1}{2 \alpha} (\eta^\mu (Q'-A)_\mu^a)^2+{\bar \psi}' [i\gamma^\mu \partial_\mu -m +gT^a\gamma^\mu Q'^a_\mu] \psi'  ]}.
\label{cfqcd1vba}
\eea
This is because a change of integration variable from unprimed variable to primed variable does not change the value of the
integration.

Using eqs. (\ref{dqa}), (\ref{psa}) and (\ref{tej}) in eq. (\ref{cfqcd1vba}) we find
\bea
&& <0|Q_\mu^a(x_1) Q_\nu^b(x_2)|0>_A=~\int [dQ] [d{\bar \psi}] [d \psi ] ~[e^{gM(x_1)}]_{ac}Q_\mu^c(x_1)~[e^{gM(x_2)}]_{bd}Q_\nu^d(x_2)\nonumber \\
 && \times e^{i\int d^4x [-\frac{1}{4}{F^a}_{\mu \nu}^2[Q] -\frac{1}{2 \alpha} (\eta^\mu ([e^{gM(x)}]_{ab}Q_\mu^b(x)))^2+{\bar \psi} [i\gamma^\mu \partial_\mu -m +gT^a\gamma^\mu Q^a_\mu] \psi ]}
\label{cfqcd1cai}
\eea
which gives
\bea
&& <0|Q_\mu^a(x_1) Q_\nu^b(x_2)|0>_A=~\int [dQ] [d{\bar \psi}] [d \psi ] ~[e^{gM(x_1)}]_{ac}Q_\mu^c(x_1)~[e^{gM(x_2)}]_{bd}Q_\nu^d(x_2)\nonumber \\
 && \times e^{i\int d^4x [-\frac{1}{4}{F^a}_{\mu \nu}^2[Q] -\frac{1}{2 \alpha} (\eta^\mu Q_\mu^a)^2+{\bar \psi} [i\gamma^\mu \partial_\mu -m +gT^a\gamma^\mu Q^a_\mu] \psi ]}.
\label{cfq5p1a}
\eea

Using the similar technique as above we find
\bea
&&
<0|[e^{gM(x_1)}]_{ac}Q_\mu^c(x_1) [e^{gM(x_2)}]_{bd}Q_\nu^d(x_2)|0>_A
 =\int [dQ] [d{\bar \psi}] [d \psi ] ~Q_\mu^a(x_1) Q_\nu^b(x_2)\nonumber \\
 && \times
e^{i\int d^4x [-\frac{1}{4}{F^a}_{\mu \nu}^2[Q] -\frac{1}{2 \alpha}(\eta^\mu Q_\mu^a)^2+{\bar \psi} [i\gamma^\mu \partial_\mu -m +gT^a\gamma^\mu Q^a_\mu] \psi  ]}
\label{cfq5p2a}
\eea
in general axial gauges
in the presence of SU(3) pure gauge background field $A^{\mu a}(x)$ as given by eq. (\ref{gtqcd}) where $M_{ab}(x)$
is given by eq. (\ref{mab}).

From eqs. (\ref{cfq5a}) and (\ref{cfq5p2a}) we find
\bea
<0|Q_\mu^a(x_1)Q_\nu^b(x_2)|0> =<0|[e^{gM(x_1)}]_{ac}Q_\mu^c(x_1) [e^{gM(x_2)}]_{bd}Q_\nu^d(x_2)|0>_A
\label{finalya}
\eea
in general axial gauges.
From eqs. (\ref{mab}), (\ref{adjin}) and (\ref{finalya}) we find
\bea
<0|Q_\mu^a(x_1)Q_\nu^b(x_2)|0> =<0|\Phi_{ac}(x_1)Q_\mu^c(x_1) \Phi_{bd}(x_2)Q_\nu^d(x_2)|0>_A
\label{finalza}
\eea
which proves factorization of soft-collinear divergences at all order in coupling constant in QCD in general axial gauges where
the non-abelian gauge link or non-abelian phase $\Phi_{ab}(x)$ in the adjoint representation of SU(3) is given by eq. (\ref{wilabf}).

\section{ Non-perturbative gluon correlation function and proof of factorization theorem in light-cone gauge }

The light-cone gauge corresponds to \cite{leib,meis,collinsp}
\bea
\eta \cdot Q^a=0,~~~~~~~~~~~~~~~~\eta^2=0
\label{jnk}
\eea
which is already covered by eqs. (\ref{zfqa}) and (\ref{axial}) where the corresponding gauge fixing term
is given by -$\frac{1}{2 \alpha} (\eta_\mu Q^{\mu a})^2$. In the light-cone coordinate system
the light-cone gauge \cite{collinsp}
\bea
Q^{+a}=0
\label{lightc}
\eea
corresponds to
\bea
\eta^\mu = (\eta^+,\eta^-,\eta_\perp)= (0,1,0)
\label{ligtc}
\eea
which covers $\eta \cdot Q^a=0$ and $\eta^2=0$ situation in eq. (\ref{jnk}).

Since eq. (\ref{finalya}) is valid for general axial gauges it is also valid for $\eta^2=0$ as given by eq. (\ref{axial}). Hence
we find from eq.  (\ref{finalya}) that
\bea
<0|Q_\mu^a(x_1)Q_\nu^b(x_2)|0> =<0|[e^{gM(x_1)}]_{ac}Q_\mu^c(x_1) [e^{gM(x_2)}]_{bd}Q_\nu^d(x_2)|0>_A
\label{finalyl}
\eea
in light-cone gauge. From eqs. (\ref{mab}), (\ref{adjin}) and (\ref{finalyl}) we find
\bea
<0|Q_\mu^a(x_1)Q_\nu^b(x_2)|0> =<0|\Phi_{ac}(x_1)Q_\mu^c(x_1) \Phi_{bd}(x_2)Q_\nu^d(x_2)|0>_A
\label{finalzl}
\eea
which proves factorization of soft-collinear divergences at all order in coupling constant in QCD in light-cone gauge where
the non-abelian gauge link or non-abelian phase $\Phi_{ab}(x)$ in the adjoint representation of SU(3) is given by eq. (\ref{wilabf}).

\section{ Non-perturbative gluon correlation function and proof of factorization theorem in general non-covariant gauges }

In QCD the generating functional with general non-covariant gauge fixing is given by \cite{noncov,noncov1}
\bea
&& Z[J,\eta,{\bar \eta}]=\int [dQ] [d{\bar \psi}] [d \psi ] ~{\rm det}(\frac{\delta (\frac{\eta^\mu \eta^\nu}{\eta^2}\partial_\mu Q_\nu^a)}{\delta \omega^b}) \nonumber \\
&& \times e^{i\int d^4x [-\frac{1}{4}{F^a}_{\mu \nu}^2[Q] -\frac{1}{2 \alpha} (\frac{\eta^\mu \eta^\nu}{\eta^2}\partial_\mu Q_\nu^a)^2+{\bar \psi} [i\gamma^\mu \partial_\mu -m +gT^a\gamma^\mu Q^a_\mu]  \psi + J \cdot Q +{\bar \psi} \eta +{\bar \eta} \psi ]}
\label{zfqn}
\eea
where $\eta^\mu$ is an arbitrary but constant four vector.
The non-perturbative gluon correlation function of the type $<0|Q_\mu^a(x_1) Q_\nu^b(x_2)|0>$ in QCD in general non-covariant gauges
is given by
\bea
&&<0|Q_\mu^a(x_1) Q_\nu^b(x_2)|0>=\int [dQ] [d{\bar \psi}] [d \psi ] ~Q_\mu^a(x_1) Q_\nu^b(x_2)\nonumber \\
&& \times ~{\rm det}(\frac{\delta (\frac{\eta^\mu \eta^\nu}{\eta^2}\partial_\mu Q_\nu^a)}{\delta \omega^b})~
e^{i\int d^4x [-\frac{1}{4}{F^a}_{\mu \nu}^2[Q] -\frac{1}{2 \alpha} (\frac{\eta^\mu \eta^\nu}{\eta^2}\partial_\mu Q_\nu^a)^2+{\bar \psi} [i\gamma^\mu \partial_\mu -m +gT^a\gamma^\mu Q^a_\mu] \psi  ]}.
\label{cfq5n}
\eea

The generating functional in the background field method of QCD with general non-covariant gauge fixing is given by \cite{noncov,noncov1}
\bea
&& Z[A,J,\eta,{\bar \eta}]=\int [dQ] [d{\bar \psi}] [d \psi ] ~{\rm det}(\frac{\delta {\cal G}^a(Q)}{\delta \omega^b}) \nonumber \\
&& \times e^{i\int d^4x [-\frac{1}{4}{F^a}_{\mu \nu}^2[A+Q] -\frac{1}{2 \alpha}
({\cal G}^a(Q))^2+{\bar \psi} [i\gamma^\mu \partial_\mu -m +gT^a\gamma^\mu (A+Q)^a_\mu] \psi + J \cdot Q +{\bar \eta} \psi +{\bar \psi} \eta  ]}
\label{azaqcdn}
\eea
where
\bea
{\cal G}^a(Q) =\frac{\eta^\mu \eta^\nu}{\eta^2} ~(\partial_\mu Q_\nu^a + gf^{abc} A_\mu^b Q_\nu^c)=\frac{\eta^\mu \eta^\nu}{\eta^2}~D_\mu[A]Q_\nu^a.
\label{gan}
\eea

The non-perturbative gluon correlation function of the type $<0|Q_\mu^a(x_1) Q_\nu^b(x_2)|0>_A$
in the background field method of QCD in general non-covariant gauges is given by
\bea
&&<0|Q_\mu^a(x_1) Q_\nu^b(x_2)|0>_A=\int [dQ] [d{\bar \psi}] [d \psi ] ~Q_\mu^a(x_1) Q_\nu^b(x_2)\nonumber \\
&& \times {\rm det}(\frac{\delta {\cal G}^a(Q)}{\delta \omega^b}) e^{i\int d^4x [-\frac{1}{4}{F^a}_{\mu \nu}^2[A+Q] -\frac{1}{2 \alpha}
({\cal G}^a(Q))^2+{\bar \psi} [i\gamma^\mu \partial_\mu -m +gT^a\gamma^\mu (A+Q)^a_\mu] \psi   ]}.
\label{cfqcdn}
\eea

By changing the integration variable $Q \rightarrow Q-A$ in the right hand side of eq. (\ref{cfqcdn}) we find
\bea
&& <0|Q_\mu^a(x_1) Q_\nu^b(x_2)|0>_A= \int [dQ] [d{\bar \psi}] [d \psi ]  ~[Q_\mu^a(x_1)- A_\mu^a(x_1) ] [ Q_\nu^b(x_2)- A_\nu^b(x_2)] \nonumber \\
&& \times {\rm det}(\frac{\delta {\cal G}_f^a(Q)}{\delta \omega^b})~~
e^{i\int d^4x [-\frac{1}{4}{F^a}_{\mu \nu}^2[Q] -\frac{1}{2 \alpha} ({\cal G}_f^a(Q))^2+{\bar \psi} [i\gamma^\mu \partial_\mu -m +gT^a\gamma^\mu Q^a_\mu] \psi  ]}
\label{cfqcd1n}
\eea
where from eq. (\ref{gan}) we find
\bea
&&{\cal G}_f^a(Q) =\frac{\eta^\mu \eta^\nu}{\eta^2}~(\partial_\mu Q_\nu^a + gf^{abc} A_\mu^b Q_\nu^c - \partial_\mu A_\nu^a)-\frac{1}{\eta^2}~gf^{abc} (\eta \cdot A^b) (\eta \cdot A^c) \nonumber \\
&& =\frac{\eta^\mu \eta^\nu}{\eta^2}~(D_\mu[A] Q_\nu^a) - \frac{\eta^\mu \eta_\nu}{\eta^2}~\partial_\mu A_\nu^a.
\label{gfan}
\eea

Changing the integration variable from unprimed variable to primed variable we find from eq. (\ref{cfqcd1n})
\bea
&& <0|Q_\mu^a(x_1) Q_\nu^b(x_2)|0>_A= \int [dQ'] [d{\bar \psi}'] [d \psi' ]~[Q_\mu'^a(x_1)- A_\mu^a(x_1) ] [ Q_\nu'^b(x_2)- A_\nu^b(x_2)] \nonumber \\
&& \times {\rm det}(\frac{\delta {\cal G}_f^a(Q')}{\delta \omega^b})~~
e^{i\int d^4x [-\frac{1}{4}{F^a}_{\mu \nu}^2[Q'] -\frac{1}{2 \alpha} ({\cal G}_f^a(Q'))^2+{\bar \psi}' [i\gamma^\mu \partial_\mu -m +gT^a\gamma^\mu Q'^a_\mu] \psi'  ]}.
\label{cfqcd1vbn}
\eea
This is because a change of integration variable from unprimed variable to primed variable does not change the value of the
integration.

Using eqs. (\ref{dqa}), (\ref{psa}) and (\ref{tej}) in eq. (\ref{cfqcd1vbn}) we find
\bea
&& <0|Q_\mu^a(x_1) Q_\nu^b(x_2)|0>_A=~\int [dQ] [d{\bar \psi}] [d \psi ] ~[e^{gM(x_1)}]_{ac}Q_\mu^c(x_1)~[e^{gM(x_2)}]_{bd}Q_\nu^d(x_2)\nonumber \\
 && \times {\rm det}(\frac{\delta {\cal G}_f^a(Q')}{\delta \omega^b})
~e^{i\int d^4x [-\frac{1}{4}{F^a}_{\mu \nu}^2[Q] -\frac{1}{2 \alpha} ({\cal G}_f^a(Q'))^2+{\bar \psi} [i\gamma^\mu \partial_\mu -m +gT^a\gamma^\mu Q^a_\mu] \psi ]}.
\label{cfqcd1cn}
\eea

From eq. (\ref{gfan}) we find
\bea
{\cal G}_f^a(Q') =
\frac{\eta^\mu \eta^\nu}{\eta^2}[\partial_\mu Q'^a_\nu + gf^{abc} A_\mu^b Q'^c_\nu] - \frac{\eta^\mu \eta^\nu}{\eta^2}~\partial_\mu A^a_\nu.
\label{gfapn}
\eea
By using eqs. (\ref{teq}) and (\ref{pg4}) in eq. (\ref{gfapn}) we find
\bea
&&{\cal G}_f^a(Q') =\frac{\eta^\mu \eta^\nu}{\eta^2}[\partial_\mu [[e^{gM(x)}]_{ab}Q_\nu^b(x) ~+ ~[\frac{e^{gM(x)}-1}{gM(x)}]_{ab}~[\partial_\nu \omega^b(x)]]\nonumber \\
&&+ gf^{abc} [\partial_\mu \omega^e(x)\left[\frac{e^{gM(x)}-1}{gM(x)}\right]_{be}] [[e^{gM(x)}]_{cd}Q_\nu^d(x) ~+ ~[\frac{e^{gM(x)}-1}{gM(x)}]_{cd}~[\partial_\nu \omega^d(x)]]]\nonumber \\
&&-\frac{\eta^\mu \eta^\nu}{\eta^2} \partial_\mu [\partial_\nu \omega^b(x)\left[\frac{e^{gM(x)}-1}{gM(x)}\right]_{ab}]
\label{gfapan}
\eea
which gives
\bea
&&{\cal G}_f^a(Q') =\frac{\eta^\mu \eta^\nu}{\eta^2}[\partial_\mu [[e^{gM(x)}]_{ab}Q_\nu^b(x)]\nonumber \\
&&+ gf^{abc} [\partial_\mu \omega^e(x)\left[\frac{e^{gM(x)}-1}{gM(x)}\right]_{be}] [[e^{gM(x)}]_{cd}Q_\nu^d(x) ~+ ~[\frac{e^{gM(x)}-1}{gM(x)}]_{cd}~[\partial_\nu \omega^d(x)]]].
\label{gfapbn}
\eea
From eq. (\ref{gfapbn}) we find
\bea
&&{\cal G}_f^a(Q') =\frac{\eta^\mu \eta^\nu}{\eta^2}[\partial_\mu [[e^{gM(x)}]_{ab}Q_\nu^b(x)]\nonumber \\
&&+ gf^{abc} [[\partial_\mu \omega^e(x)]\left[\frac{e^{gM(x)}-1}{gM(x)}\right]_{be}] [[e^{gM(x)}]_{cd}Q_\nu^d(x)]]
\label{gfapcn}
\eea
which gives
\bea
&&{\cal G}_f^a(Q') =\frac{\eta^\mu \eta^\nu}{\eta^2} [[e^{gM(x)}]_{ab}\partial_\mu Q_\nu^b(x)+Q_\mu^b(x)\partial_\nu [[e^{gM(x)}]_{ab}]\nonumber \\
&&+  [[\partial_\mu \omega^e(x)]\left[\frac{e^{gM(x)}-1}{gM(x)}\right]_{be}]gf^{abc} [[e^{gM(x)}]_{cd}Q_\nu^d(x)]].
\label{gfapdn}
\eea
Using eq. (\ref{pg4k}) in (\ref{gfapdn}) we find
\bea
&&{\cal G}_f^a(Q') = \frac{\eta^\mu \eta^\nu}{\eta^2}[e^{gM(x)}]_{ab}\partial_\mu Q_\nu^b(x)
\label{gfapen}
\eea
which gives
\bea
({\cal G}_f^a(Q'))^2 = (\frac{\eta^\mu \eta^\nu}{\eta^2}\partial_\mu Q_\nu^a(x))^2.
\label{gfapfn}
\eea
From eqs. (\ref{detr}) and (\ref{gfapen}) we find
\bea
&&{\rm det} [\frac{\delta {\cal G}_f^a(Q')}{\delta \omega^b}] ={\rm det}
[\frac{\eta^\mu \eta^\nu}{\eta^2}\frac{ \delta [[e^{gM(x)}]_{ac}\partial_\mu Q_\nu^c(x)]}{\delta \omega^b}]={\rm det}[\frac{\eta^\mu \eta^\nu}{\eta^2}
[e^{gM(x)}]_{ac}\frac{ \delta (\partial_\mu Q_\nu^c(x))}{\delta \omega^b}]\nonumber \\
&&=\left[{\rm det}[
[e^{gM(x)}]_{ac}]\right]~\left[{\rm det}[\frac{\eta^\mu \eta^\nu}{\eta^2}\frac{ \delta (\partial_\mu Q_\nu^c(x))}{\delta \omega^b}]\right]={\rm exp}[{\rm Tr}({\rm ln}[e^{gM(x)}])]~{\rm det}[\frac{\eta^\mu \eta^\nu}{\eta^2}\frac{ \delta (\partial_\mu Q_\nu^a(x))}{\delta \omega^b}]\nonumber \\
&&={\rm det}[\frac{\eta^\mu \eta^\nu}{\eta^2}\frac{ \delta (\partial_\mu Q_\nu^a(x))}{\delta \omega^b}].
\label{gqp4an}
\eea
Using eqs. (\ref{gfapfn}) and (\ref{gqp4an}) in eq. (\ref{cfqcd1cn}) we find
\bea
&&<0|Q_\mu^a(x_1) Q_\nu^b(x_2)|0>_A=\int [dQ] [d{\bar \psi}] [d \psi ] ~[e^{gM(x_1)}]_{ac}Q_\mu^c(x_1)~[e^{gM(x_2)}]_{bd}Q_\nu^d(x_2)\nonumber \\
 && \times
{\rm det}(\frac{\eta^\mu \eta^\nu}{\eta^2}\frac{\delta (\partial_\mu Q_\nu^a)}{\delta \omega^b})~
e^{i\int d^4x [-\frac{1}{4}{F^a}_{\mu \nu}^2[Q] -\frac{1}{2 \alpha}(\frac{\eta^\mu \eta^\nu}{\eta^2}\partial_\mu Q_\nu^a)^2+{\bar \psi} [i\gamma^\mu \partial_\mu -m +gT^a\gamma^\mu Q^a_\mu] \psi  ]}.
\label{cfq5p1n}
\eea
Using the similar technique as above we find
\bea
&&
<0|[e^{gM(x_1)}]_{ac}Q_\mu^c(x_1) [e^{gM(x_2)}]_{bd}Q_\nu^d(x_2)|0>_A
 =\int [dQ] [d{\bar \psi}] [d \psi ] ~Q_\mu^a(x_1) Q_\nu^b(x_2)\nonumber \\
 && \times {\rm det}(\frac{\eta^\mu \eta^\nu}{\eta^2}\frac{\delta (\partial_\mu Q_\nu^a)}{\delta \omega^b})~
e^{i\int d^4x [-\frac{1}{4}{F^a}_{\mu \nu}^2[Q] -\frac{1}{2 \alpha}(\frac{\eta^\mu \eta^\nu}{\eta^2}\partial_\mu Q_\nu^a)^2+{\bar \psi} [i\gamma^\mu \partial_\mu -m +gT^a\gamma^\mu Q^a_\mu] \psi  ]}
\label{cfq5p2n}
\eea
in general non-covariant gauges
in the presence of SU(3) pure gauge background field $A^{\mu a}(x)$ as given by eq. (\ref{gtqcd}) where $M_{ab}(x)$
is given by eq. (\ref{mab}).

From eqs. (\ref{cfq5n}) and (\ref{cfq5p2n}) we find
\bea
<0|Q_\mu^a(x_1)Q_\nu^b(x_2)|0> =<0|[e^{gM(x_1)}]_{ac}Q_\mu^c(x_1) [e^{gM(x_2)}]_{bd}Q_\nu^d(x_2)|0>_A
\label{finalyn}
\eea
in general non-covariant gauges.
From eqs. (\ref{mab}), (\ref{adjin}) and (\ref{finalyn}) we find
\bea
<0|Q_\mu^a(x_1)Q_\nu^b(x_2)|0> =<0|\Phi_{ac}(x_1)Q_\mu^c(x_1) \Phi_{bd}(x_2)Q_\nu^d(x_2)|0>_A
\label{finalzn}
\eea
which proves factorization of soft-collinear divergences at all order in coupling constant in QCD in general non-covariant gauges where
the non-abelian gauge link or non-abelian phase $\Phi_{ab}(x)$ in the adjoint representation of SU(3) is given by eq. (\ref{wilabf}).

\section{ Non-perturbative gluon correlation function and proof of factorization theorem in general Coulomb gauge }

In QCD the generating functional with general Coulomb gauge fixing is given by \cite{noncov,noncov1}
\bea
&& Z[J,\eta,{\bar \eta}]=\int [dQ] [d{\bar \psi}] [d \psi ] ~{\rm det}(\frac{\delta ([g^{\mu \nu}-\frac{n^\mu n^\nu}{n^2}]\partial_\mu Q_\nu^a)}{\delta \omega^b}) \nonumber \\
&& \times e^{i\int d^4x [-\frac{1}{4}{F^a}_{\mu \nu}^2[Q] -\frac{1}{2 \alpha} ([g^{\mu \nu}-\frac{n^\mu n^\nu}{n^2}]\partial_\mu Q_\nu^a)^2+{\bar \psi} [i\gamma^\mu \partial_\mu -m +gT^a\gamma^\mu Q^a_\mu]  \psi + J \cdot Q +{\bar \psi} \eta +{\bar \eta} \psi ]}
\label{zfqc}
\eea
where
\bea
n^\mu =(1,0,0,0).
\eea
The non-perturbative gluon correlation function of the type $<0|Q_\mu^a(x_1) Q_\nu^b(x_2)|0>$ in QCD in general Coulomb gauge
is given by
\bea
&&<0|Q_\mu^a(x_1) Q_\nu^b(x_2)|0>=\int [dQ] [d{\bar \psi}] [d \psi ] ~Q_\mu^a(x_1) Q_\nu^b(x_2)\nonumber \\
&& \times ~{\rm det}(\frac{\delta ([g^{\mu \nu}-\frac{n^\mu n^\nu}{n^2}]\partial_\mu Q_\nu^a)}{\delta \omega^b})~
e^{i\int d^4x [-\frac{1}{4}{F^a}_{\mu \nu}^2[Q] -\frac{1}{2 \alpha} ([g^{\mu \nu}-\frac{n^\mu n^\nu}{n^2}]\partial_\mu Q_\nu^a)^2+{\bar \psi} [i\gamma^\mu \partial_\mu -m +gT^a\gamma^\mu Q^a_\mu] \psi  ]}.\nonumber \\
\label{cfq5c}
\eea
The generating functional in the background field method of QCD with general Coulomb gauge fixing is given by \cite{noncov,noncov1}
\bea
&& Z[A,J,\eta,{\bar \eta}]=\int [dQ] [d{\bar \psi}] [d \psi ] ~{\rm det}(\frac{\delta {\cal G}^a(Q)}{\delta \omega^b}) \nonumber \\
&& \times e^{i\int d^4x [-\frac{1}{4}{F^a}_{\mu \nu}^2[A+Q] -\frac{1}{2 \alpha}
({\cal G}^a(Q))^2+{\bar \psi} [i\gamma^\mu \partial_\mu -m +gT^a\gamma^\mu (A+Q)^a_\mu] \psi + J \cdot Q +{\bar \eta} \psi +{\bar \psi} \eta  ]}
\label{azaqcdc}
\eea
where
\bea
{\cal G}^a(Q) = [g^{\mu \nu}-\frac{n^\mu n^\nu}{n^2}](\partial_\mu Q_\nu^a + gf^{abc} A_\mu^b Q_\nu^c)=[g^{\mu \nu}-\frac{n^\mu n^\nu}{n^2}]D_\mu[A]Q_\nu^a.
\label{gac}
\eea
The non-perturbative gluon correlation function of the type $<0|Q_\mu^a(x_1) Q_\nu^b(x_2)|0>_A$
in the background field method of QCD in general Coulomb gauge is given by
\bea
&&<0|Q_\mu^a(x_1) Q_\nu^b(x_2)|0>_A=\int [dQ] [d{\bar \psi}] [d \psi ] ~Q_\mu^a(x_1) Q_\nu^b(x_2)\nonumber \\
&& \times {\rm det}(\frac{\delta {\cal G}^a(Q)}{\delta \omega^b}) e^{i\int d^4x [-\frac{1}{4}{F^a}_{\mu \nu}^2[A+Q] -\frac{1}{2 \alpha}
({\cal G}^a(Q))^2+{\bar \psi} [i\gamma^\mu \partial_\mu -m +gT^a\gamma^\mu (A+Q)^a_\mu] \psi   ]}.
\label{cfqcdc}
\eea
Hence by replacing $\frac{\eta^\mu \eta^\nu}{\eta^2} \rightarrow [g^{\mu \nu}-\frac{n^\mu n^\nu}{n^2}]$ everywhere in the derivations
in the previous section we find
\bea
<0|Q_\mu^a(x_1)Q_\nu^b(x_2)|0> =<0|\Phi_{ac}(x_1)Q_\mu^c(x_1) \Phi_{bd}(x_2)Q_\nu^d(x_2)|0>_A
\label{finalzc}
\eea
which proves factorization of soft-collinear divergences at all order in coupling constant in QCD in general Coulomb gauge where
the non-abelian gauge link or non-abelian phase $\Phi_{ab}(x)$ in the adjoint representation of SU(3) is given by eq. (\ref{wilabf}).

\section{ Correct Definition of the Gluon Distribution Function at High Energy Colliders }

Under the non-abelian gauge transformation as given by
eq. (\ref{aftgrmpi}) we find from eqs. (\ref{ttt}), (\ref{hqg}), (\ref{adjin}) and (\ref{wilabf})
that the the non-abelian gauge gauge link or non-abelian phase in QCD in the adjoint representation
of SU(3) transforms as
\bea
\Phi'_{ab}(x)= \left[e^{gM(x)}\Phi(x)\right]_{ab},~~~~~~~~~~~~~~~~~~M_{ab}(x)=f^{abc}\omega^c(x).
\label{wilabfadj}
\eea
Hence from eqs. (\ref{jpr}) and (\ref{wilabfadj}) we find that
$<0|\Phi_{ac}(x_1)Q_\mu^c(x_1) \Phi_{bd}(x_2)Q_\nu^d(x_2)|0>_A$ in eq. (\ref{finalz}) [or in (\ref{finalza})
or in (\ref{finalzl}) or in eq. (\ref{finalzn}) or in eq. (\ref{finalzc})]
is gauge invariant and eq. (\ref{finalz}) [or (\ref{finalza}) or (\ref{finalzl})
or eq. (\ref{finalzn}) or eq. (\ref{finalzc})] is consistent with the factorization of soft-collinear
divergences at all order in coupling constant in QCD.

Hence from eqs. (\ref{finalz}) [or (\ref{finalza}) or (\ref{finalzl})
or (\ref{finalzn}) or (\ref{finalzc})] and (\ref{wilabf}) we find that the correct definition of the
gluon distribution function at high energy colliders which is
consistent with the number operator interpretation of the gluon and is gauge invariant and is consistent with the
factorization theorem in QCD is given by
\bea
&& f_{g/P}(x)= \frac{P^+}{x2\pi}\int dy^- e^{-ix{P}^+ y^- }\nonumber \\
&&\times \left[<P| [{\cal P}e^{-ig\int_0^{\infty} d\lambda~ l \cdot { A}^c(y + \lambda l )T^{(A)c}}] Q_\mu^a (y)
[{\cal P}e^{-ig\int_0^{\infty} d\lambda~ l \cdot { A}^c(\lambda l )T^{(A)c}}]Q^{\mu a}(0)|P>\right]_{y^+=y_T=0}\nonumber \\
\label{gpdffpio2}
\eea
which is valid in covariant gauge, in light-cone gauge, in general axial gauges, in general non-covariant gauges and in
general Coulomb gauge etc. respectively. Eq. (\ref{gpdffpio2}) can be written as
\bea
&& f_{g/P}(x)= \frac{P^+}{x2\pi}\int dy^- e^{-ix{P}^+ y^- } \times <P| Q_\mu^a (0,y^-,0_T)
[{\cal P}e^{ig\int_0^{(0,y^-,0_T)} dz \cdot { A}^c(z )T^{(A)c}}]Q^{\mu a}(0)|P>\nonumber \\
\label{gpdffpiof}
\eea
which reproduces eq. (\ref{gpdffpi}). This completes the derivation of the correct definition of the
gluon distribution function at high energy colliders from first principles.

\section{Conclusions}
Unlike QED, since $F_{\mu \nu}^a(x)F^{\mu \nu a}(0)$ in QCD contains cubic and quartic powers of the gluon field the present
definition of the gluon distribution function at high energy colliders is not consistent with the number operator
interpretation of the gluon. In this paper we have derived the correct definition of the gluon distribution function
at high energy colliders from first principles which is consistent with the number operator interpretation of the gluon and is gauge
invariant and is consistent with the factorization theorem in QCD. After renormalization the gluon distribution function
is expected to obey a QCD evolution equation, like DGLAP equation \cite{tung}, which follows from renormalization group
equation.

\acknowledgments

I thank J. C. Collins for useful discussions, in particular, with respect to absence of rapidity divergence in transverse momentum
integrated parton distribution function when light-like Wilson line is used in QCD. I also thank D. E. Soper for useful discussions.

\end{document}